\documentclass[10pt,twocolumn,preprintnumbers,reprint,amsmath,amssymb,amsfonts,aps,prd,superscriptaddress,showpacs,nofootinbib]{revtex4-1}
\usepackage{graphicx}
\usepackage{hyperref}
\usepackage[normalem]{ulem}
\usepackage{slashed}
\usepackage{float}
\usepackage[caption=false]{subfig}
\usepackage{bm}
\usepackage[utf8]{inputenc}
\usepackage{color}
\usepackage[table,xcdraw,dvipsnames]{xcolor}
\hypersetup{
	bookmarks=true,         
	unicode=true,          
	pdftoolbar=true,        
	pdfmenubar=true,        
	pdffitwindow=false,     
	pdfstartview={FitH},    
	pdftitle={iDM at SBN},    
	pdfauthor={Author},     
	pdfsubject={Subject},   
	pdfcreator={Creator},   
	pdfproducer={Producer}, 
	pdfkeywords={keyword1} {key2} {key3}, 
	pdfnewwindow=true,      
	colorlinks=true,       
	linkcolor=black,          
	citecolor=black,        
	filecolor=black,      
	urlcolor=black           
}

\newcommand{\gfour}{{\sc GEANT4}}
\newcommand{\bdnmc}{{\sc BdNMC}}
\newcommand{\md}{{\sc MadDump}}
\newcommand{\amc}{{\sc MadGraph5}\_a{\sc MC@NLO}}
\newcommand{\genie}{{\sc GENIE}}
\newcommand{\pythia}{{\sc Pythia8}}

\def\mdm{M_{\chi_1}}
\def\mdp{ M_{V}}
\def\mhds{M_{\chi_2}}
\def\dchi{\Delta_{\chi}{}}
\def\Rx{R_{\chi}}

\def\gmtwomu{(g-2)_\mu}
\newcommand*{\INFNFR}{Istituto Nazionale di Fisica Nucleare, Laboratori Nazionali di Frascati, C.P. 13, 00044 Frascati, Italy}

\begin{document}
	
\title{Inelastic Dark Matter at the Fermilab Short Baseline Neutrino Program}

\preprint{PITT-PACC-2111}

\author{Brian Batell}
\affiliation{PITT PACC, Department of Physics and Astronomy, University of Pittsburgh, Pittsburgh, PA 15260, USA}
\author{Joshua Berger}
\affiliation{Department of Physics, Colorado State University, Fort Collins, CO 80523, USA}
\author{Luc Darm\'e}
\affiliation{\INFNFR}
\author {Claudia Frugiuele} 
\affiliation{INFN, Sezione di Milano, Via Celoria 16, I-20133 Milano, Italy}

\newcommand{\LD}[1]{{\color{Bittersweet}{#1}}}
\newcommand{\JB}[1]{{\color{red}{#1}}}
\newcommand{\BB}[1]{{\color{blue}{#1}}}
\newcommand{\CF}[1]{{\color{green}{#1}}}

	\begin{abstract}
	We study the sensitivity of the Fermilab Short-Baseline Neutrino (SBN) experiments, MicroBooNE, ICARUS, and SBND, to MeV- to GeV-scale inelastic dark matter interacting through a dark photon mediator. These models provide interesting scenarios of light thermal dark matter, which, while challenging to probe with direct and indirect detection experiments, are amenable to accelerator-based searches. We consider production of the dark sector states with both the Fermilab Booster $8$ GeV and NuMI $120$ GeV proton beams and study the signatures of scattering and decay of the heavy excited dark state in the SBN detectors. These distinct signatures probe complementary regions of parameter space. All three experiments will be able to cover new ground, with an excellent near-term opportunity to search for cosmologically motivated targets explaining the observed dark matter abundance.
	\end{abstract}
		\vspace{1cm}
		\maketitle
	\tableofcontents
	\setcounter{footnote}{0}

\section{Introduction}

A host of disparate gravitational phenomena provides seemingly incontrovertible evidence for dark matter (DM), yet it is striking how little we know about its fundamental properties. This stark situation has spurred extensive explorations for novel theories and new experimental probes of DM over the past decade. While many interesting ideas have emerged, the hypothesis of a light dark sector weakly coupled to the Standard Model through a portal interaction is particularly compelling. Viable models of thermally produced light DM below the Lee-Weinberg bound~\cite{Lee:1977ua} can be realised in this framework~\cite{Boehm:2003hm,Pospelov:2007mp,Feng:2008ya}. %
Experiments at the intensity frontier have a vital role to play in probing the low masses and feeble couplings characteristic of such theories. In particular, proton beam fixed-target experiments, including accelerator neutrino beam experiments, offer an interesting testing ground for new light, weakly coupled particles and DM~\cite{Batell:2009di,Essig:2010gu,deNiverville:2011it,deNiverville:2012ij,Morrissey:2014yma,Batell:2014yra,Soper:2014ska,Dobrescu:2014ita,Kahn:2014sra,Gardner:2015wea,Coloma:2015pih,Izaguirre:2017bqb,Frugiuele:2017zvx,Darme:2017glc,Jordan:2018gcd,Berlin:2018pwi,deNiverville:2018hrc,deNiverville:2018dbu,Darme:2018jmx,Buonocore:2018xjk,DeRomeri:2019kic,Dutta:2019nbn,Berlin:2020uwy,Dutta:2020vop,Breitbach:2021gvv}. 
Due to the large collision luminosity and forward lab-frame kinematics, a substantial forward flux of relativistic dark sector states may be produced in the primary proton-target collisions. These dark particles may subsequently be observed through their decays or scatterings as they traverse a detector located downstream of the target. 
These experiments provide a critical component of a broader experimental program of searches at medium-energy $e^+e^-$ colliders, fixed-target missing energy/momentum experiments, low mass direct detection experiments, rare particle decay experiments, and the LHC, which will provide powerful sensitivity to light dark sectors and DM in the MeV-GeV mass range~\cite{Alexander:2016aln,Battaglieri:2017aum,Beacham:2019nyx,BRNreport,Lanfranchi:2020crw}.

A particularly promising near-term venue for proton fixed target dark sector searches is provided by the Fermilab Short-Baseline Neutrino (SBN) program. The SBN program comprises three liquid argon time projection chamber (LArTPC) detectors -- 
SBND~\cite{Antonello:2015lea}, MicroBooNE~\cite{Acciarri:2016smi}, and ICARUS~\cite{Amerio:2004ze} -- 
installed along the Booster Neutrino Beam (BNB)~\cite{Machado:2019oxb}. 
Interestingly, dark sector particles can be produced both from the medium energy (8 GeV) on-axis
Booster proton beam and from the higher energy (120 GeV) Main Injector NuMI proton beam.
Concerning the latter, MicroBooNE and ICARUS are located approximately 8$^\circ$ and 6$^\circ$ off-axis
with respect to the NuMI beam, implying that the associated flux of dark particles passing through
these detectors can be substantial.
Several studies have highlighted the exciting sensitivity of the SBN experiments to light exotic new particles, including long-lived heavy neutral leptons~\cite{Ballett:2016opr}, DM tridents~\cite{deGouvea:2018cfv}, Higgs portal scalars~\cite{Batell:2019nwo}, and elastically scattering DM~\cite{Buonocore:2019esg}, and the first experimental searches have been carried out by MicroBooNE~\cite{Abratenko:2019kez,Abratenko:2021ebm}. 
With SBND coming on-line soon and ICARUS now operating, 
significant improvements in reach are expected over the next several years. 

In this work we investigate the near-term prospects for the SBN experiments to probe models of MeV-GeV scale inelastic dark matter (iDM) coupled to the SM through a kinetically mixed dark photon. First proposed in the context of the DAMA anomaly~\cite{TuckerSmith:2001hy,Bernabei:2013xsa}, iDM models as advocated in~\cite{Izaguirre:2015zva,Izaguirre:2017bqb} have received broader attention in recent years  as a compelling scenario for sub-GeV thermal DM~\cite{Jordan:2018gcd,Chang:2018rso,DAgnolo:2018wcn,Berlin:2018pwi,Berlin:2018bsc,Darme:2018jmx,Berlin:2018jbm,Pospelov:2019vuf,Tsai:2019mtm,Duerr:2019dmv,Jodlowski:2019ycu,Duerr:2020muu,Kim:2016zjx,Giudice:2017zke,Kim:2018veo,Chatterjee:2018mej,Ha:2018obm}. The off-diagonal nature of the DM coupling to a heavier excited state provides a simple mechanism to evade otherwise stringent bounds from direct detection experiments and cosmic microwave background observations. On the other hand, simple iDM models also feature new potential signatures associated with DM produced in accelerator experiments, including inelastic up- and down-scattering and semi-visible decays of the excited state to DM and a pair of charged leptons. While the sensitivity of accelerator-based neutrino experiments to iDM has been studied in other contexts~\cite{Izaguirre:2017bqb,Jordan:2018gcd}, here we consider the specific experimental issues related to their production in the Booster and NuMI beam lines and their detection in the SBN LArTPC detectors. In particular, this is to our knowledge the first time that the sensitivities of the three upcoming SBN experiments to iDM scenarios have been estimated. Furthermore, we have for the first time singled out and quantified the neutrino-induced background relevant for long-lived decay for experiments with angular reconstruction capability, as well as estimated the background rejection efficiency of angular cuts. In particular, searches for on-axis dark sector production from the Booster at SBND and off-axis production from NuMI at ICARUS can explore significant new regions of parameter space, including those predicting the observed DM relic density. 

The paper is structured as follows. In Section~\ref{sec:model}
 we present the iDM model to be studied in this work, including a discussion of the long-lived excited dark state and the DM relic abundance. Our modelling of the iDM signal and neutrino-induced backgrounds at the SBN experiments, as well as our analysis strategy, is discussed in Section~\ref{sec:analysis}. Our projections for the SBN sensitivity to the iDM model are presented in in Section~\ref{sec:results}, and we provide our conclusions and outlook in Section~\ref{sec:conclusion}.

\section{Light inelastic dark matter}
\label{sec:model}

In this section we discuss a simplified model of sub-GeV iDM interacting via a dark photon mediator that can be probed by Fermilab SBN experiments. After introducing the model, we discuss the thermal relic abundance predictions and the existing experimental constraints on this scenario.

\subsection{Model}

We consider a model of fermionic iDM based on a spontaneously broken $U(1)_D$ gauge symmetry with a massive dark photon mediator, $V_\mu$~\cite{Izaguirre:2015zva}.\footnote{See Refs.~\cite{Cui:2009xq,Batell:2009vb} for earlier studies of the same model in the context of weak scale iDM.} 
The dark photon interacts with the SM via kinetic mixing with the ordinary photon. 
The Lagrangian describing the dark photon is
\begin{equation}
\mathcal{L} \supset -\frac{1}{4} V_{\mu\nu}V^{\mu\nu} +\frac{1}{2} \mdp^2\, V_\mu V^{\mu} +  \frac{\varepsilon}{2} V_{\mu\nu}F^{\mu\nu} \ ,
\end{equation}
where $M_{V}$ is the dark photon mass, $\varepsilon$ is the kinetic mixing strength, and $F_{\mu\nu}$ is the SM photon field strength. 
The kinetic mixing generates a coupling of the dark photon to the electrically charged particles with strength suppressed by $\varepsilon$. 
In the physical basis, the dark photon interactions with the SM are thus described by the coupling $\varepsilon \, e V_\mu J_{EM}^\mu$,
where $J_{EM}^\mu$ is the SM electromagnetic current. 

The dark matter sector consists of two Majorana fermion states $\chi_1$ and $\chi_2$, with Lagrangian
\begin{equation}
\mathcal{L} \supset  \sum_{i = 1,2} \ \frac{1}{2} \overline \chi_i \, i \gamma^\mu  \,\partial_\mu \chi_i
-\frac{1}{2} M_{\chi_i}\, \overline \chi_i \chi_i \, .
\end{equation}
Here $M_{\chi_i}$ denotes the physical fermion masses, and we assume $M_{\chi_1} < M_{\chi_2}$ such that $\chi_1$ is the stable DM candidate. 
Two important parameters that will appear in our phenomenological considerations are the fractional mass splitting $\Delta_\chi$, defined as
 \begin{equation}
 \Delta_\chi  ~\equiv~ \frac{M_{\chi_2} - M_{\chi_1}}{M_{\chi_1}},
 \end{equation}
 and the dark photon-to-DM mass ratio $R_\chi$:
 \begin{align}
    \Rx ~\equiv~ \frac{\mdp}{\mdm}\ .
\end{align}
It is clear that $\mdm$, $\dchi$ and $\Rx$ parameterise the spectrum of the dark sector.
The dark sector particles are assumed to dominantly couple to the dark photon mediator through an off-diagonal interaction,
\begin{equation}
\mathcal{L} \supset  i \,g_D \, V_\mu \, \overline \chi_2 \gamma^\mu \chi_1,
\label{eq:Vchi2chi1}
\end{equation}
where $g_D \equiv \sqrt{4 \pi \alpha_D}$ is the $U(1)_D$ gauge coupling.  

The simplified model presented here can be realised as the low energy limit of a renormalisable completion involving a dark Higgs field with a $U(1)_D$ charge of 2 and a Dirac fermion field with $U(1)_D$ charge of 1. Gauge and Yukawa couplings of the dark Higgs field generate both a dark photon mass term and small Majorana masses for the Weyl components of the Dirac fermion after spontaneous $U(1)_D$ breaking. Provided the Dirac mass is much larger than these Majorana masses, the dominant interaction of the DM is given by equation~(\ref{eq:Vchi2chi1}). Diagonal couplings of the form $V_\mu \, \overline \chi_i \gamma^\mu \gamma^5 \chi_i$ are suppressed by the ratio of the difference in Majorana masses to the Dirac mass. Due to an enhanced charged conjugation symmetry, it is technically natural for the Majorana mass difference to be small or even vanishing~\cite{Berlin:2018jbm}. We note that the dark Higgs boson cannot in general be completely decoupled from the spectrum at large $\alpha_D \sim 0.5$ while maintaining perturbativity~\cite{Darme:2017glc}. However, as long as its mass remains of the order of the dark photon mass it does not significantly affect the phenomenology~\cite{Darme:2018jmx}, and thus it will thus not be considered in this work.

The iDM scenario enjoys several attractive features. First, for $M_{\chi_1} +M_{\chi_2}  \lesssim M_V$, the observed DM relic abundance can be obtained through thermal freeze-out of DM co-annihilation to SM particles for certain choices of model parameters, providing predictive cosmologically motivated targets for experiments.
This will be discussed in detail below in Section~\ref{sec:relic-density}. 
Second, the heavier Majorana fermion state $\chi_2$ decays rapidly for $\Delta_\chi M_{\chi_1}> 2 m_e$, implying that DM annihilation is strongly suppressed for temperatures below the mass splitting. Thus, stringent constraints on late-time DM annihilation from precision studies of the cosmic microwave background temperature anisotropies do not apply to this scenario~\cite{Ade:2015xua,Slatyer:2009yq}.
Lastly, the non-relativistic scattering of DM off of SM particles in direct detection experiments, $\chi_1 \, {\rm SM} \rightarrow \chi_2 \, {\rm SM}$, faces a strong kinematic suppression even for very small mass splittings due to the inelastic (endothermic) nature of the transition\footnote{See Refs.~\cite{Finkbeiner:2009mi,Batell:2009vb,Graham:2010ca,Harigaya:2020ckz,Lee:2020wmh,Baryakhtar:2020rwy,Bramante:2020zos,Bloch:2020uzh} for studies of direct detection via exothermic transitions.}.
Therefore, accelerator experiments can provide a unique probe of this simple and well-motivated sub-GeV DM scenario.

\subsection{Relic density for iDM}
\label{sec:relic-density}

The underlying mechanisms relevant for the cosmological production of light iDM have been thoroughly studied in the literature.
Our goal here is to obtain a simple recasting procedure of the relic density targets obtained from the full evolution of the Boltzmann equations from Refs.~\cite{Izaguirre:2017bqb,Berlin:2018bsc,Duerr:2019dmv} for generic values of $\dchi$ and $\Rx$. 
The relic density arises from $\chi_1 \chi_2$ co-annihilation to SM particles, which can be resonantly enhanced when $M_V \sim M_{\chi_1} +  M_{\chi_2}$. 
This requires a modification of the standard instantaneous freeze-out approximation~\cite{Griest:1990kh}. In particular, we will build our recasting procedure in the resonant region following the approach developed in~\cite{Feng:2017drg}.

The effective thermally-averaged cross-section takes the form:
\begin{align}
\label{eq:sigmaeff}
    \langle \sigma v\rangle_{\rm eff} =  \frac{ 2 (1+\dchi)^{3/2} \, e^{-x\dchi}}{(1+(1+\dchi)^{3/2} \, e^{-x \dchi})^2} \times  \ (\sigma v)_0 ,
\end{align}
where we have included the number of degrees of freedom of each DM component, and the second term reads:
\begin{align}
    (\sigma v)_0 = \begin{cases} (\sigma v)_{\rm lab}^{s = M_V^2} \frac{2 \sqrt{\pi}\Gamma_V M_V}{(\mdm + \mhds)^2} \sqrt{\epsilon_R} x^{3/2} e^{-x\epsilon_R} \ \textrm{for } \ \epsilon_R \ll1,   \\[0.5em]
 (\sigma v)_{\rm lab}^{v=0} \quad \textrm{otherwise}, \\\end{cases}
\end{align}
with $\epsilon_R = \frac{M_V^2}{(\mdm + \mhds)^2}-1$ and $(\sigma v)_{\rm lab}$ the total DM annihilation cross-section times velocity to any SM states $\chi_1 \bar \chi_2 \to \rm SM$. 
The first case corresponds to the resonant limit with $\Gamma_V \ll M_V$ where most of the annihilation occurs directly on the resonance, while the second is the standard result for a $s$-wave cross-section. In order to include all the potential hadronic channels, we use the R-ratio approach~\cite{Izaguirre:2015zva}, defining
\begin{align}
    R(s)= \frac{\sigma (e^+ e^- \to \rm hadrons)}{\sigma (e^+ e^- \to \mu \mu)},
\end{align}
which we have extracted from \texttt{DarkCast}~\cite{Ilten:2018crw,Tanabashi:2018oca}. The final relic density scales as:
\begin{align}
\label{eq:intOmh}
    \Omega h^2 \propto\left( \sqrt{g_{\rm eff} (x_f)} \int_{x_f}^\infty \frac{\langle \sigma v\rangle_{\rm eff} }{x^2} dx\right)^{-1} \ ,
\end{align}
where $g_{\rm eff}$ is the SM effective number of degrees of freedom evaluated at the freeze-out temperature $x_f$. We can estimate $x_f$ following the standard equation~\cite{Griest:1990kh}:
\begin{align}
\label{eq:FOxf}
    x_f = \log \left\{\! \frac{0.038 m_{\rm pl} \, \mdm \langle \sigma v\rangle_{\rm eff}}{ \sqrt{x_f \, g_{\rm eff} }}\, 2 \left[1\! + \! (1 \! + \! \dchi)^{3/2} \, e^{-x_f \dchi} \right] \! \right\},
\end{align}
with $m_{\rm pl} = 1.22 \times 10^{19} \, \rm GeV$ the Planck mass and $g_{\rm eff}$ the SM effective number of degrees of freedom. Note that as shown in Eq.~\eqref{eq:sigmaeff}, $\langle \sigma v\rangle_{\rm eff}$ contains another exponential factor of $ x_f$. Based on the above ingredients, we can obtain a simple estimate of the relic density. Since numerous recent works have presented full solutions of the Boltzmann equations for different choices of parameters~\cite{Izaguirre:2017bqb,Berlin:2018bsc,Duerr:2019dmv}, we will instead use the ingredients above to interpolate between the full relic density lines of~\cite{Duerr:2019dmv}.\footnote{ We use the relic density targets for the parameters sets $ (\dchi=0.4, \Rx=2.5)$, $ (\dchi=0.4, \Rx=3)$, $ (\dchi=0, \Rx=2)$ and $ (\dchi=0.1, \Rx=2.5)$.} In more details, the steps are the following. 

\begin{itemize}
      \item Solve  Eqs.~\eqref{eq:intOmh} and~\eqref{eq:FOxf} for both the reference relic density line and for the new values $\dchi$ and $\Rx$, leading to $ (\Omega h^2)^{\rm ref}$, $(\Omega h^2)^{\rm new}$ , and to the ratio:
    \begin{align}
        r_{\rm ref} \, \equiv \, \frac{(\Omega h^2)^{\rm ref}}{0.12} \sim \mathcal{O}(1)\ ,
    \end{align}
    where we used the latest DM relic density estimate from the Planck collaboration~\cite{Aghanim:2018eyx}, and since our estimate is based on the instantaneous freeze-out the ratio is in general not precisely $1$, particularly for large $\dchi$ and strongly resonant cross-sections.
    \item Determine the final relic density target in $\varepsilon_f$ from solving 
        \begin{align}
        (\Omega h^2)^{\rm new} = 0.12 \times r_{\rm ref}.
    \end{align}
 \end{itemize}
In practice, the above procedure is thus simply a ``smart'' interpolation between the existing full calculation results using the instantaneous freeze-out results. We have cross-checked our interpolations with the independent results from Ref.~\cite{Berlin:2018bsc}.

\subsection{Existing constraints}
\label{sec:otherlimits}

A large variety of intensity frontier experiments have the potential to search for iDM-like signatures. Analyses from past experiments have been re-interpreted in recent years, and new results from modern experiments have also been presented. We list below the most important ones along with the relevant limits we include in our final result.

\paragraph{CHARM} The CERN-Hamburg-Amsterdam-Rome-Moscow experiment, in addition to its main on-axis detector, used an off-axis calorimeter module to search for new long-lived particles, including heavy neutral leptons~\cite{Santoni:1986yw,Bergsma:1983rt} and axion-like particles~\cite{Bergsma:1985qz}. These limits have been re-interpreted in the context of various dark sector models in the literature.
In particular, the limits from CHARM on the iDM scenario have recently been considered in~\cite{Tsai:2019mtm}. 
Note that we extend the previous re-interpretation by further including the direct parton-level production, improving the CHARM limit for the large mass regime.  We have simulated parton-level production of dark sector states using the same strategy as for the SBN program described in Sec.~\ref{sec:signalgeneration}.

Although the original search from CHARM relied on observing two charged particles, the analysis did not resolve each track. 
Instead they required that the observed pulse height in a scintillator plate, placed directly in front of the main calorimeter, should  be larger than $1.5$ minimum ionizing particles (corresponding to $9$ MeV). Coupled with the requirement that no more than 2 tracks should be seen in the calorimeter, these cuts lead to no detected events. 

\paragraph{NuCal} Using the $70$ GeV proton beam from the U70 accelerator, NuCal accumulated  $2\cdot 10^{18}$ protons on target (PoT)~\cite{Barabash:1992uq,Blumlein:1990ay}. Similarly to CHARM, it featured a large decay volume and could therefore operate in a near-zero background mode. We use the limits based on~\cite{Tsai:2019mtm}.

\paragraph{LSND} Using the Los Alamos Meson Physics Facility $0.8$ GeV proton beam, the LSND experiment accumulated $\sim 10^{23}$ PoT~\cite{Athanassopoulos:1996ds}. 
A measurement of neutrino-electron elastic scattering analysed in~\cite{Athanassopoulos:1996ds,Auerbach:2001wg} has been recasted to place limits on the iDM model, and we use the limits derived in~\cite{Izaguirre:2017bqb} and~\cite{Tsai:2019mtm} depending on the model parameter sets used.  It is worth mentioning that this result was derived from the reinterpretation of neutrino scattering results, assuming that the $e^+ e^-$ pair would be reconstructed as a single electron for small enough angular separation with a similar efficiency.

\paragraph{NA64} The NA64 experiment, a fixed-target experiment employing a CERN SPS secondary 100 GeV electron beam, has accumulated $2.84 \times 10^{11}$ electron-on-target (EoT), with plans for collecting $5 \times 10^{12}$ EoT in the coming years. Based on an active target, it searches for large missing energy due to a dark photon escaping the target~\cite{NA64:2019imj}. It mostly constrains the low mass region.

\paragraph{BaBar} 
Based on the $e^+ e^-$ PEP-II B-factory, the BaBar experiment has accumulated a total luminosity of 53 fb$^{-1}$, which was analysed for single photon events with large missing energy~\cite{Lees:2017lec}. This analysis places strong constraints on dark photons that decay invisibly to DM particles. However, for iDM with large mass splittings, the heavy state $\chi_2$ instead decays semi-visibly within the detector, which significantly weakens the bounds for large dark photon masses~\cite{Mohlabeng:2019vrz,Duerr:2019dmv}. 
\paragraph{Experiments sensitive to DM scattering} The MiniBooNE-DM collaboration has recently used the MiniBooNE detector in an ``off-target'' run configuration to search for light elastic sub-GeV DM~\cite{Aguilar-Arevalo:2017mqx,Aguilar-Arevalo:2018wea} as described in more detailed below. For large dark gauge coupling $\alpha_D \sim 0.5$, the corresponding limit is competitive with NA64 and we will include it when relevant.
We also note that re-interpretations of analyses from other past experiments sensitive to DM-electron scattering, such as BEBC~\cite{Grassler:1986vr,Buonocore:2019esg}, E137~\cite{Bjorken:1988as,Batell:2014mga,Marsicano:2018glj,Marsicano:2018krp}, NOvA~\cite{Bian:2017axs,deNiverville:2018dbu}, can have competitive sensitivity. 
See Ref.~\cite{Buonocore:2019esg} for a recent study and comparison.

\section{Searching for iDM at the SBN program}
\label{sec:analysis}

\begin{table*}[t]
	\centering
		\begin{tabular}{|c|c|c|c| c| c | }
		\hline
				\rule{0pt}{14pt}Experiment & $E^{\rm kin}_{\rm beam}$&  Target & PoT  & D (m) &  $W \times H \times L $ (m${}^3$)    \\[0.2em]
				\hline
				\hline
				\rule{0pt}{14pt}
			 MicroBooNE~\cite{Antonello:2015lea,Acciarri:2016smi}  & &   &   &$470$ & $ 2.6 \times 2.3 \times 10.4 $    \\
			 SBND~\cite{Antonello:2015lea} & \ $8$ GeV \ & $1.7\lambda$ Be& \ $6.6 \cdot 10^{20}$ \  &$110$ &  $ 4 \times 4 \times 5$    \\
			 ICARUS~\cite{Yu:2020noc} &  &  &   &$600$ & $6.0 \times 3.2 \times 18.0 $     \\
			\hline
					\rule{0pt}{14pt}
			 MicroBooNE Off-target & &  &   &$420$ & $ 2.6 \times 2.3 \times 10.4 $    \\
			 SBND Off-target& $8$ GeV& \ Thick Fe \ & $2.2 \cdot 10^{20}$  &$60$ &  $ 4 \times 4 \times 5$    \\
			 ICARUS Off-target & &   &   &$550$ & $6.0 \times 3.2 \times 18.0 $     \\
				\hline
			\rule{0pt}{14pt}
    	ICARUS OA~\cite{Buonocore:2019esg}& $120$ GeV & $2\lambda$ C & $7.7 \cdot 10^{21}$ &$785$ m, $\theta = 0.1$  & $6.0 \times 3.2 \times 18.0 $  \\
	    	\hline
			\end{tabular}
		\caption{Beam and target characteristics for various experiments, along with the anticipated total number of protons on target (PoT). OA indicates ''off-axis'' and $\lambda$ is the interaction length of the corresponding material. The distance (to the centre of the experiment) $D$ and typical detector length $L$ are further indicated. Note that the geometry used is the active volume, around $25-30 \%$ larger than the fiducial volume.
		}
		\label{tab:exp} 
\end{table*}

The SBN experiments at Fermilab include the Short-Baseline Near Detector (SBND), MicroBooNE, and ICARUS. These detectors all employ liquid argon time projection chambers (LArTPCs) to detect tracks of charged particles. The detectors are placed along the Booster Neutrino Beam (BNB), which is generated by striking a  1.7 nuclear interaction length beryllium target with 8 GeV kinetic energy protons. 
The BNB also has the capability to run in ``off-target'' mode, during which the proton beam is steered past the beryllium target and directed onto the iron absorber at the end of the decay volume, $50$ m downstream of the target. In off-target mode the beam-related neutrino flux is significantly reduced compared to the normal running mode, allowing for enhanced sensitivity to new physics signals. This off-target run configuration was used by the MiniBooNE-DM collaboration to search for light elastic sub-GeV DM~\cite{Aguilar-Arevalo:2017mqx,Aguilar-Arevalo:2018wea}. While there are currently no plans to perform an off-target run with the SBN experiments, we will investigate the potential sensitivity of such a run to the iDM model below

In addition to this beam, the ICARUS detector is less than 6 degrees off-axis from the Neutrinos at the Main Injector (NuMI) beam, generated by 120 GeV protons impacting a 2-interaction-length graphite target.  
We thus also consider the sensitivity of the ICARUS detector to off-axis production of iDM in the NuMI beam (ICARUS-OA) 
We have summarised the various experimental setups considered in this work in Table~\ref{tab:exp}.

\subsection{iDM signatures in neutrino experiments} 
\label{sec:kinthre}

Dark sector particles interacting through the vector portal are typically produced in proton beam dump experiments through the following three different mechanisms \cite{deNiverville:2016rqh}: 
\begin{itemize}
\item {\it Pseudoscalar meson decay.} Low mass dark photons may be efficiently produced through the decays $\pi^0, \eta, \eta' \rightarrow \gamma V$.
\item {\it Proton bremsstrahlung.}
Dark photons can be directly produced via bremsstrahlung, $pp \rightarrow p V X$, which dominates for $M_V \sim 1$ GeV~\cite{Blumlein:2013cua,deNiverville:2016rqh}
\item {\it Drell-Yan.} For higher beam energies and masses, dark sector particles can be directly produced via the parton-level process  $q \bar{q} \rightarrow V \rightarrow \chi_1 \chi_2$.
\end{itemize}

The iDM model leads to several potential signatures in proton fixed-target experiments including up-scattering, down-scattering, and semi-visible $\chi_2$ decay ~\cite{Izaguirre:2017bqb}.
Which signature dominates is primarily dictated by the lifetime of the excited state. 
\begin{itemize}
\item {\it Long-lived.} If the lifetime of $\chi_2$ is long enough for it to reach the detector, it may then either down-scatter to a DM state $\chi_1$ or decay to a DM particle $\chi_1$ and visible SM particles through an off-shell dark photon. In this case the scattering signatures (both $ \chi_1 e \to \chi_2 e$ and $\chi_2 e \to \chi_1 e$) would give rise to a subdominant contribution with respect to the decay signatures.
\item {\it Quasi-stable.} In the region where the $\chi_2$ lifetime is too long, the dominant signatures are $ \chi_1 e \to \chi_2 e$ and $\chi_2 e \to \chi_1 e$, which both mimic the standard DM elastic scattering signature.
\item {\it Short-lived.} If $\chi_2$ is too short lived to reach the detector the only signature is the up-scattering $ \chi_1 e \to \chi_2 e$. Depending on the $\chi_2$ lifetime, $\chi_2$ can
decay back into the neutrino detector, leading to a displaced vertex signature. Otherwise the up-scattering mimics the standard DM elastic scattering.

\end{itemize}
Hence there are three distinct regions corresponding to three characteristic phenomenological signatures: 
(i) up-scattering followed by decay $\chi_2$ in the detector, 
(ii) decay of beam-produced $\chi_2$ in the detector, 
and (iii) scattering.
In order to understand which signatures dominate in different regions of the parameter space we need to study the $\chi_2$ decay width.
In the regime $M_V \gg M_1 \gg  \Delta_\chi M_1 > 2 m_e$, the excited state decays via $\chi_2 \rightarrow \chi_1 e^+ e^-$. 
The partial decay width for this process is given by~\cite{Izaguirre:2015zva} 
\begin{equation}
\label{eq:decaywidth}
\Gamma_{\chi_2 \rightarrow \chi_1 e^+ e^-} \simeq \frac{4 \varepsilon^2 \alpha \alpha_D \Delta_\chi^5 M_{\chi_1}^5}{15 \pi M_{V}^4} \ .
\end{equation}
The strong dependence on the small mass splitting leads to macroscopic decay lengths even for relatively large couplings. For the experiments of interest here, the lab frame decay length is 
\begin{align}
\label{eq:lifetime}
\gamma\beta c \tau_{\chi_2} & \approx 10^3 \, {\rm m} \, 
\left(\frac{\varepsilon}{10^{-3}}\right)^{-2} 
    \left(\frac{\alpha_D}{0.1}\right)^{-1}
\left(\frac{\Delta_\chi}{0.1}\right)^{-5} \nonumber \\
&~~~~ \times \left(\frac{M_{\chi_1}}{200\,{\rm MeV}}\right)^{-1}
\left(\frac{M_V}{3M_{\chi_1}}\right)^4
 \left(\frac{\gamma\beta c}{10}\right) .
\end{align}
 
\begin{figure}[t!]
	\includegraphics[width=0.99\linewidth]{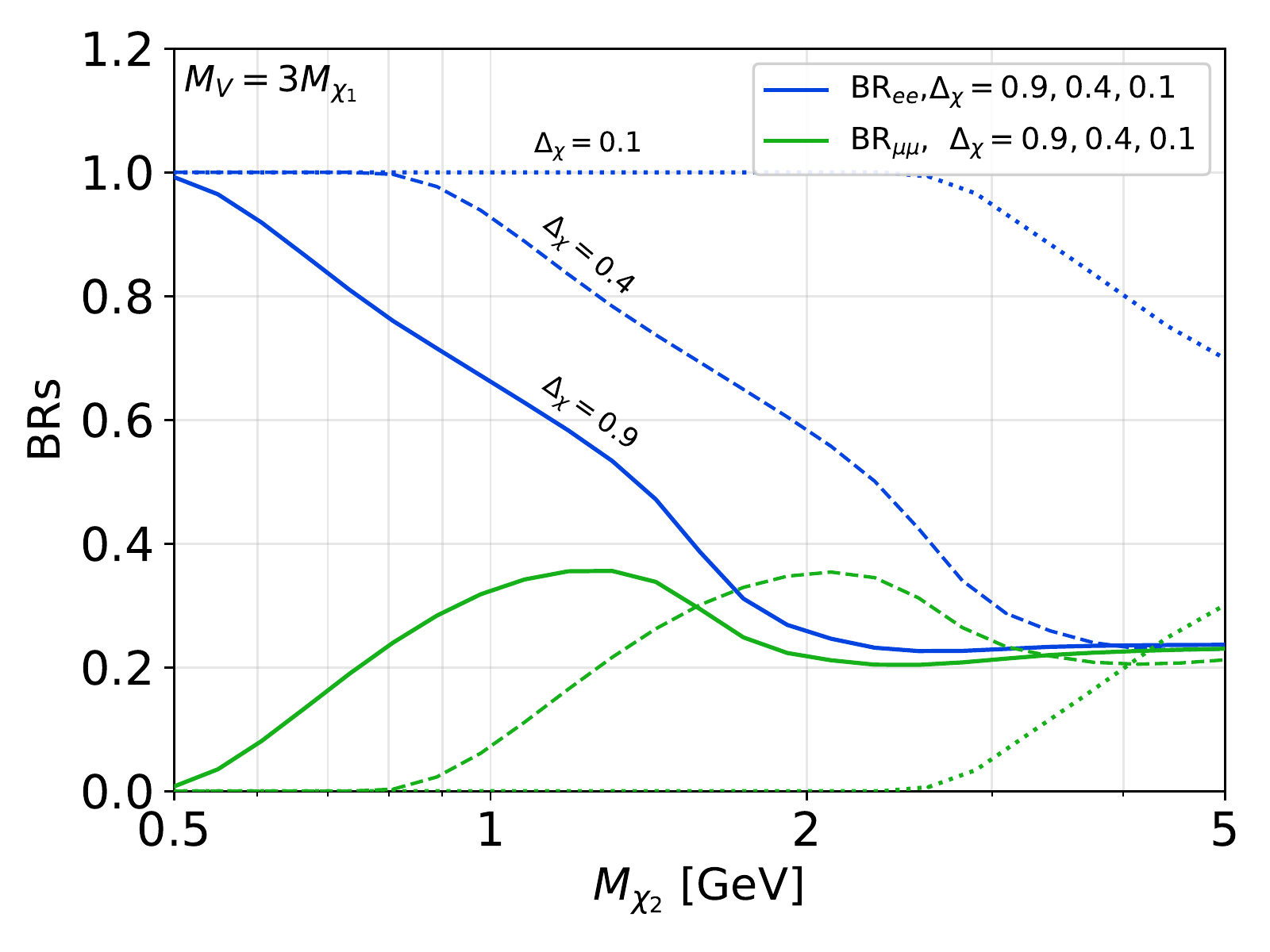}
	\vspace{-0.5cm}
\caption{Leptonic branching ratios of $\chi_2$ as function of its mass for various values of the splitting $\dchi=0.1,0.4,0.9$.}
	\label{fig:BRV}
\end{figure}

In order to fully discuss the iDM phenomenology, a comment regarding the decay modes of $\chi_2$ is necessary to assess both the pure decay and the upscattering followed by decay phenomenology.
For mass splittings less than the dimuon threshold, $\Delta_\chi M_{\chi_1} < 2 M_\mu$, the excited state decays via
$\chi_2 \rightarrow \chi_1 e^+ e^-$.
For larger $\chi_1-\chi_2$ mass splittings, new leptonic and hadronic channels open up. We have included these additional channels based on the $R$-ratio approach, obtaining the differential hadronic decay rate $\chi_2 \rightarrow \chi_1 + \rm (hadrons)$ via:
\begin{align}
    \frac{d  \Gamma^{\rm had.}_{\chi_2}}{d s } =     \frac{d  \Gamma^{\rm \mu \mu}_{\chi_2}}{d s } R(s) \ ,
\end{align}
where $\sqrt{s}$ is the momentum of the off-shell dark photon (as done in e.g.~\cite{Jodlowski:2019ycu}). We present the leptonic branching ratios for several values of the splitting parameter $\Delta_\chi$ in Fig.~\ref{fig:BRV}. In particular, we find that the branching ratio suppression for the $e^+ e^-$ final state is always fairly mild, reduced by at most a factor of $3$ for the largest masses accessible in beam dump experiments. We note that in the opposite regime of very small splittings  below the di-electron threshold, $\Delta_\chi M_{\chi_1} < 2 m_e$, the excited state decays via $\chi_2 \rightarrow \chi_1+ 3 \gamma$ and is stable on cosmological time scales~\cite{Finkbeiner:2009mi,Batell:2009vb}.

Overall, Eq.~(\ref{eq:lifetime}) provides insight into the characteristic phenomenology for given parameter choices, as we now discuss in further detail.

\paragraph{Long-lived decay $\chi_2 \to \chi_1 e^+ e^-$}

For small splittings, small kinetic mixing and intermediate range masses the dominant signature consists of a beam-produced excited state $\chi_2$, which is long lived enough to travel to the detector and decay semi-visibly.
A particularly important aspect of the considered three-body decay $\chi_2 \to \chi_1 e^+ e^-$ is that the angle of the final $e^+ e^-$ is correlated with the energy of the pair and the mass splitting $\dchi$. More precisely, since the total exchanged squared momentum in the off-shell dark photon $s$ satisfies $s < (\dchi M_{\chi_1})^2$  we can decompose $s$ in the lab frame in terms of the energy and angle of the $e^+ e^-$ pair, leading to \cite{Izaguirre:2017bqb}
\begin{align}
\label{eq:MassThreshold}
    \mdm^2  \Delta_\chi^2 > 2 m_e^2 (1 + \cos \theta_c) + 2 E_c^2 (1 - \cos \theta_c) \ ,
\end{align}
where, in the laboratory frame, we used an angular cut $\theta_c$ on the $e^+ e^-$ angle and and energy cut $E_c$ on both $e^+$ and $e^-$.
For small angular separation and a typical energy threshold $E_c = 30$ MeV, we easily get an approximation for the lower accessible mass threshold:
\begin{align}
\label{eq:MassThresholdSimp}
\mdm \gtrsim \frac{ \sqrt{1 + 860 \, \theta_c^2} }{\dchi} \, \rm MeV .
\end{align}

\paragraph{Down-scattering and up-scattering signatures}

We now turn to the region of very small $\dchi$ and small $M_{\chi_1}$, for which both up-scattering and down-scattering signatures resembles those of regular elastic scattering since the $\chi_2$ is quasi-stable on detector length scales. In this parameter region our focus will be on scattering with electrons, which yields a clean forward energetic single electron signature. The cross section for both up- and down-scattering is the same as for standard elastic scattering up to small corrections of order $\dchi \mdm/E$ and $\dchi \mdm/m_V$~\cite{Izaguirre:2017bqb}. The approximate differential cross section with respect to the outgoing electron energy $E_f$ in the laboratory frame is~\cite{Batell:2014mga}  
\begin{align}
\label{eq:csscat}
\frac{d \sigma_{\rm scat}}{d E_f} =  4 \pi \varepsilon^2 \alpha \alpha_D \frac{2 m_e E^2 - f(E_f)(E_f - m_e)}{(E^2 - m_{\chi_1}^2)(m_{V}^2 + 2 m_e E_f - 2 m_e^2)^2}, 
\end{align}
where $E$ is the incoming $\chi_1$ or $\chi_2$ energy and $f(E_f) = 2 m_e E -m_e E_f + m_\chi^2 + 2 m_e^2$.

\paragraph{Up-scatter followed by decay}
Finally, for the large $\dchi$ and large $\varepsilon$ region, the main signature is $\chi_1$ up-scattering followed by a prompt or displaced $\chi_2$ decay. We focus on $\chi_1$-nucleon up-scattering in the large $\dchi$ region for a couple of reasons. First, up-scattering off electrons is kinematically suppressed for large mass splittings, particularly for DM produced with the lower energy BNB. The $\chi_1$ energy threshold $E_{\rm min}$ to initiate an up-scattering on a target particle of mass $m$ is 
 \begin{align}
 \label{eq:upscatter-threshold}
E \geq E_{\rm min} &= \frac{M_{\chi_2}^2 - M_{\chi_1}^2 
+ 2 m M_{\chi_2} }{2m} \\
& \simeq \frac{M_{\chi_1}^2 \Delta_{\chi} (2+ \Delta_{\chi})}{2 m} ~~~~ {\rm for} ~ m \ll M_{\chi_{1,2}}. \nonumber
 \end{align}
We see that for large splittings the typical energy for up-scattering off electrons is much greater than for scattering off nucleons. 
 Second, while elastic DM-nucleon scattering searches typically must contend with large neutrino-induced backgrounds~\cite{Aguilar-Arevalo:2017mqx,Aguilar-Arevalo:2018wea}, for the case of iDM each up-scattering process can lead to an associated visible decay vertex, providing an additional handle to reject the backgrounds and leverage the large DM-nucleon scattering rate. 
 We have included both the above electron-scattering cross-section (for the small splitting regime) as well as up-scattering on nucleons via the public code \bdnmc ~\cite{deNiverville:2016rqh}, adapting the implemented cross-section to the Dirac fermion case using the results of Ref.~\cite{Batell:2014yra}. 

 As the visible charged particles from the $\chi_2$ decay are expected to constitute the main signal, we have not implemented cuts on the outgoing nucleon kinematics.
 Our projections for the expected number of events should be considered as conservative estimates, particularly for ICARUS in an off-axis configuration where the typical DM energy ranges between $5$ to $30$ GeV.  Note that we do not include the potentially relevant deep-inelastic scattering processes, but these would also be worth exploring in a full study.

\subsection{Signal simulation chain}
\label{sec:signalgeneration}

It is clear from the discussion above that an accurate description of our signal events requires a complete  simulation of  the iDM production at the target and its subsequent propagation to the SBN detectors, followed by the decay and/or scattering processes. The main steps of our simulation chain are presented below. 

For meson-induced production in the BNB, we simulate $\pi^0$ and $\eta$ meson distributions using the Sanford-Wang empirical distribution~\cite{Sanford:1967zza} using the fits made by the MiniBooNE collaboration~\cite{AguilarArevalo:2008yp} as implemented in \bdnmc~\cite{deNiverville:2016rqh}. 
When considering the BNB-OT and the NuMI-OA configurations, we rely instead on the meson distributions stored in the database of Ref.~\cite{Celentano:2020vtu,celentano_andrea_2020_3890984}. 
The latter are based on a \gfour~\cite{Brun:1994aa}~simulation of the meson production in the current NOvA target for NuMI-OA and in the iron dump used during the MiniBooNE run~\cite{AguilarArevalo:2008yp} for  BNB Off-target. 
For larger masses, we use the proton bremsstrahlung production as implemented in \bdnmc~\cite{deNiverville:2016rqh}, which includes resonant $\rho/\omega$ vector meson mixing using the time-like form factor derived from~\cite{Faessler:2009tn}. When relying on parton-level production, we instead use the \md~\cite{Buonocore:2018xjk} plugin for \amc~\cite{Alwall:2014hca} to generate the distribution of both $\chi_1$ and $\chi_2$  from an $s$-channel dark photon. 

An important subtlety is that the proton bremsstrahlung process as implemented in \bdnmc~relies on the assumption that
\begin{align}
\label{eq:bremrequirements}
    E_{\rm beam}, E_V, E_{\rm beam} - E_V \gg M_V, m_p, p^V_T ,
\end{align}
where  $E_{\rm beam}$ is the incoming proton energy, $m_p$ the proton mass, $E_V$ the outgoing dark photon energy in the lab frame, and $p_T^V$ its transverse momentum~\cite{deNiverville:2016rqh}. For the on-axis BNB beam, we use the cut $E_V/E_{\rm beam} \subset [0.3,0.7], \, p^V_T < 0.2 \, \rm GeV$. For the NuMI-OA, the situation is peculiar in that one needs $p^V_T \sim 0.1 E_V$ in order to point toward the off-axis ICARUS detector. In order to satisfy Eq.~\eqref{eq:bremrequirements}, we use 
\begin{align}
E_V/E_{\rm beam} \subset [0.05,0.7], \ p^V_T < 10\, \rm GeV .
\end{align}

All production mechanisms are then centralised to a modified version of \bdnmc~\cite{Darme:2017glc,Darme:2018jmx}, which further includes the propagation of the dark states to the detector and the decay of the heavy dark state $\chi_2 \to \chi_1 e^+ e^-$ via an off-shell dark photon. For the large splitting regime where additional decay channels open, we include them in the estimate for the $\chi_2$ lifetime and in the corresponding branching ratio suppression as we focus on $e^+ e^-$ final states in this work. At the decay stage, the final state is directly analysed and the various event selection cuts described in the next sections are applied to the Monte Carlo truth. 
When considering up-scattering and down-scattering signatures, we use the built-in routines from \bdnmc~\cite{deNiverville:2016rqh,Batell:2014yra}, adapting the cross-sections to the pseudo-Dirac DM case considered in this work as discussed above.

\subsection{Backgrounds and selection cuts}
\label{sec:bkd}

There are three sources of background for the signal we consider. Cosmic rays deposits that are not associated to the cosmic ray tagger can mimic signal or interfere with reconstruction. We do not consider this background in our analysis and assume that cosmic rays signals can be efficiently removed. We also neglect interactions of beam neutrinos in the dirt surrounding the detector hall. The third background, which we do include in our analysis is interactions of beam neutrinos in the fiducial volume. 

The dominant signature expected in the SBN experiments being a $e^+e^-$ pair arising from a $\chi_2$ decay, we focus first on the neutrino-induced background relevant for this final state. We then discuss the case of down-scattering and up-scattering before concluding with some remarks on the use of timing information.

\paragraph{$\chi_2$ decay signature and angular separation}

The $e^+e^-$ pair produced in decays of the heavy dark matter state $\chi_2 \to \chi_1 e^+ e^-$ undergo bremsstrahlung in the detector, leading to a shower of $e^+$ and $e^-$ that are detected by the TPC. Provided that the separation between the original $e^+e^-$ is sufficiently large, the individual four-momenta of these instigating particles can be reconstructed. This allows for detailed information about potential signal events to be measured and for separation from the backgrounds as we discuss below.

There are several ways in which neutrino interactions can mimic the signal of an $e^+e^-$ appearing with no other activity. These include misidentified hadrons or muons, $\gamma\gamma$ of $e\gamma$ with misidentified photons, and single $\gamma$ conversion, along with no other reconstructed activity. We assume perfect identification of hadrons and muons in our analysis.  Furthermore, we assume that $\gamma\gamma$ and $e\gamma$ can be distinguished by the increased charge of one or both of the resulting showers and the displacement of the photon shower vertices from the neutrino interaction point by one radiation length. We focus on the photon conversion background.

To study this background, we apply the following event generation scheme. The flux of beam neutrinos is taken from Ref.~\cite{Admas:2013xka}. The interaction of the beam neutrinos with argon is simulated using \genie~\cite{Andreopoulos:2009rq,Andreopoulos:2015wxa}. The unstable hadrons other than charged pions and neutrons from 
\genie~final state particles are decayed using \pythia~\cite{Sjostrand:2007gs}. Photons are injected into a very large volume of liquid argon in \gfour~\cite{Brun:1994aa}. If the first interaction they undergo is conversion, which is the dominant process at the relevant energies, the resulting $e^-e^+$ pair is added to the event. If they undergo Compton scattering instead, the resulting hard electron is added to the event. For any other type of interaction, which is an extremely rare occurrence, no new particles are added. For example, a tiny number of events have photons that undergo the photoelectric effect and those photons are ignored. The initial photon is then removed from the event.

We proceed to study the kinematics of this final set of particles. We do not apply energy or angular smearing to the particles. We veto events containing the following particles if they appear above the corresponding kinetic energy threshold: protons above $50$ MeV, charged pions and muons above $30$ MeV. This is based on Ref.~\cite{Acciarri:2015uup}, but we use a lower charged pion threshold as calorimetric reconstruction of charged pions will not be required here. Neutrons are not considered in our analysis. Events are required to have exactly two electrons above a kinetic energy of 30 MeV. We then examine the total number of background events from this procedure, as well as those that have an electron pair with an opening angle larger than $5^\circ$ and $10^\circ$. The results of these selection criteria are shown in Table~\ref{tab:bkd}. 
We further apply these kinematic criteria to the signal to determine a signal reconstruction efficiency. 
It is worth noting that electrons at an opening angle of even $5^\circ$ may not be easily reconstructed as a pair of particles. Since we apply these criteria to both the signal and background events, it is assumed that even if distinct particles are not reconstructed, an analysis can be developed to find events that look like a merged pair of showers.  Since we apply our criteria to both signal and background, we should have a good estimate of the total expected number of such events.

In our analysis, we rely only on kinematic information about the particles produced in the signal heavy state decays and the background neutrino scattering in argon. Additional information may be used in an experimental analysis that could help to further reduce the background. Neutrino scattering can lead to production of a recoiling neutron. Neutrons are rather challenging to reconstruct as they tend to bounce around the detector, leaving small localised deposits whenever they hit a nucleus. The ability to reconstruct neutrons is still under study in LArTPCs and we ignore neutrons entirely in our analysis. The ability to veto neutrons, as well as short proton tracks (i.e.\ those from protons below $50$ MeV), would help to reduced the background significantly. Furthermore, neutrino scattering produces a nuclear remnant that will decay into a stable nucleus. The de-excitation of the nuclear remnant leads to low energy photons and is challenging to detect, but is certainly a distinguishing feature of the background compared to the signal. 

\paragraph{Down-scattering and up-scattering background}

We next discuss the case of up- and down-scattering signatures. 
For the small splitting regime with standard DM-electron scattering signature, we require the scattered electron to be extremely forward in order to mitigate the beam neutrino backgrounds (see e.g.~\cite{Aguilar-Arevalo:2017mqx}), and we also impose an energy cut similar to that for the decay signature:
\begin{align}
\label{eq:CutsScatter}
    \cos \theta > 0.99,  \qquad E_{e} >\,  30 \, \rm MeV \ ,
\end{align}
where $\theta$ is the angle of the outgoing electron with respect to the interaction point direction and $E_e$ its energy in the laboratory frame. This cut leaves the neutrino-electron scattering as the dominant source of background. In our projections we will consider the electron scattering signature for both ICARUS-OA and the SBN off-target run. 
Given the small neutrino-electron scattering rate, the off-axis location of ICARUS with respect to the NuMI beamline, and the suppressed neutrino flux in off-target run mode, one can expect a near background free search in these configurations after Eq.~(\ref{eq:CutsScatter}) is imposed~\cite{Dobrescu:2014ita,Aguilar-Arevalo:2017mqx}. 

Up-scattering off electrons with subsequent decay in the detector is a distinctive signal with few relevant backgrounds. 
If the decay is prompt, then the signal effectively is three electromagnetic showers with one being at narrow angle with respect to the beam. While this signal could be challenging to reconstruct if the showers overlap, there are few backgrounds from either neutrinos or cosmic rays that could fake it. The most plausible is charged current electron neutrino interactions with inelastic $\pi^0$ production and a short-distance photon conversion. Every piece of this is rare (electron neutrinos, inelastic pion production, and short-distance photon conversion), so the background is expected to be very small. It is possible for parameters where the decay is a bit displaced with respect to the up-scatter that this $\nu_e$ CC + $\pi^0$ background could be more significant, but it is likely still too rare to pose a significant challenge. For an off-target run, this potential background would be even further reduced. It is therefore plausible that such a search would be effectively background free. 
As discussed above, up-scattering off electrons is typically kinematically suppressed in the large splitting regime where $\chi_2$ undergoes rapid decay (see Eq.~(\ref{eq:upscatter-threshold})), and for this reason we will not investigate it further in this work. However, we note that it could still be relevant for lower DM masses and the higher energy DM produced in the NuMI beam, particularly given the distinctive characteristics of the signal.

Similar signals with up-scatter off nucleons face more potential backgrounds from neutrino scatter off nucleons that are less well studied. Neutrino-nucleon scattering with additional production of $\pi^0$ in particular are a potential background for $\chi_2$ decay lengths in the tens of cm. For short $\chi_2$ lifetimes, charged-current production of $p$-$e^-$ could also pose a challenge if the electron shower is misreconstructed as two separate showers. As there are no backgrounds that exactly mimic the signal, there are many handles to enhance the signal-to-background ratio. Furthermore, an off-target run would further reduce possible neutrino backgrounds. Although a full study of reconstruction and background reduction in this scenario is beyond the scope of this work, background should not be a major impediment to this signal.
 
\paragraph{Timing}

In some parts of parameter space, it could be advantageous to exploit timing information about events. Neutrinos travel at speeds very close to the speed of light, while heavier  states may travel appreciably slower, leading to a lag relative to the backgrounds. LArTPC detectors have relatively coarse timing capabilities, with microsecond time resolution, leading to delayed signals only for very slowly moving  states. The LArTPC detectors are, however, equipped with photomultiplier tubes (PMT) used to detect scintillation light associated with charged particles passing through the liquid argon that can yield nanosecond timing resolution. This could dramatically increase the power of timing to distinguish signal from background, but is as yet untested and so we do not exploit this possibility. 

Timing information has been partially used by the MicrobooNE collaboration to study detection of heavy neutral leptons~\cite{Abratenko:2019kez}. In their search, cuts were applied to look for events that fell outside of the entire batch window of $1.5~\mu\text{s}$. This sort of timing analysis is somewhat simpler to develop, but is not suitable for more highly boosted particles and discards a significant number of signal events. On the other hand, it allows for drastically reduced backgrounds.

\begin{table}[t]
	\centering
	\begin{tabular}{c|c p{1cm} p{1cm} }
		\hline
				\rule{0pt}{14pt}
		$\theta_{\rm cut}$	&  $\qquad 0^\circ\qquad $  & $5^\circ$ & $10^\circ$ \\
	    	\hline
	    \rule{0pt}{14pt}	SBND 	&  	$3093$ &  $ 68 $& $13.8$ \\[0.2em]
	    	MicroBooNE \	&  	$87$ &   $1.8$ & $0.22$  \\[0.2em]
	    	ICARUS-OA \	&  	$218$ &  $ 4.4$ & $1.0$ \\[0.2em]
	    	\hline
			\end{tabular}
		\caption{Total background events counts for SBND, MicroBooNE and ICARUS in an off-axis configurations, the number are given for $10^{20}$ PoT.
		}
		\label{tab:bkd} 
\end{table}

\section{Results}
\label{sec:results}

In this section, we present our sensitivity projections of the SBN experiments to the iDM model. These projections are derived 
based on our simulations for iDM production, $\chi_2$ decays, and $\chi_{1,2}$ scattering, as well as our treatments of detector thresholds and backgrounds (in the case of the $\chi_2$ decay signature), as discussed in Section~\ref{sec:analysis}.
The SBN experiments will be able to probe two qualitatively distinct parameter regions that predict the observed DM relic density and are currently unconstrained: 1) DM masses in the 100 MeV - 1 GeV scale with relatively large mass splittings $\dchi \gtrsim 0.2$, and 2) DM masses in the 10 - 100 MeV scale with relatively small mass splittings $\dchi\lesssim 0.1$. The former is viable due to a sensitivity gap between beam dump and colliders searches that emerges for moderate $\chi_2$ decay lengths, $c \tau_{\chi_2} \sim {\cal O}({\rm mm} - {\rm cm})$,
while the latter corresponds to the opposite case of long-lived or quasi-stable $\chi_2$.  
We will discuss the prospects of SBN experiments in both regimes.

\subsection{Large mass splitting region}

When the mass splitting between $\chi_2$ and $\chi_1$ becomes large enough to lead to a relatively short-lived $\chi_2$, the sensitivities from beam dump experiments, such as CHARM and NuCal, is limited by the distance between their detector and the beam target. Moreover, for moderate values of $\varepsilon$, $e^+ e^-$ colliders such as BaBar have limited sensitivity due to smaller production rates and luminosity, or, in some cases, the lack of dedicated searches for displaced decays. This creates a sensitivity gap in the iDM parameter space corresponding to sub-ns $\chi_2$ lifetimes (similar to the well-known ``Mont's gap'' in visibly decaying dark photon searches).  
Based on this discussion, it is clear that the SBN experiments, with baselines of order hundreds of meters, will not have sensitivity to primary beam-produced $\chi_2$ decays in this large mass splitting parameter region. Additionally, we expect that for such large values of $\dchi$, $\chi_1$ up-scattering off electrons will typically be kinematically suppressed, particularly for iDM produced with the Booster beam. On the other hand, the SBN experiments can instead efficiently search for $\chi_1$ up-scattering off nucleons, $\chi_1 N \rightarrow \chi_2 N$, followed by the fast semi-visible decay $\chi_2 \rightarrow \chi_1 e^+ e^-$. 

\begin{figure}[t]
	\includegraphics[width=0.99\linewidth]{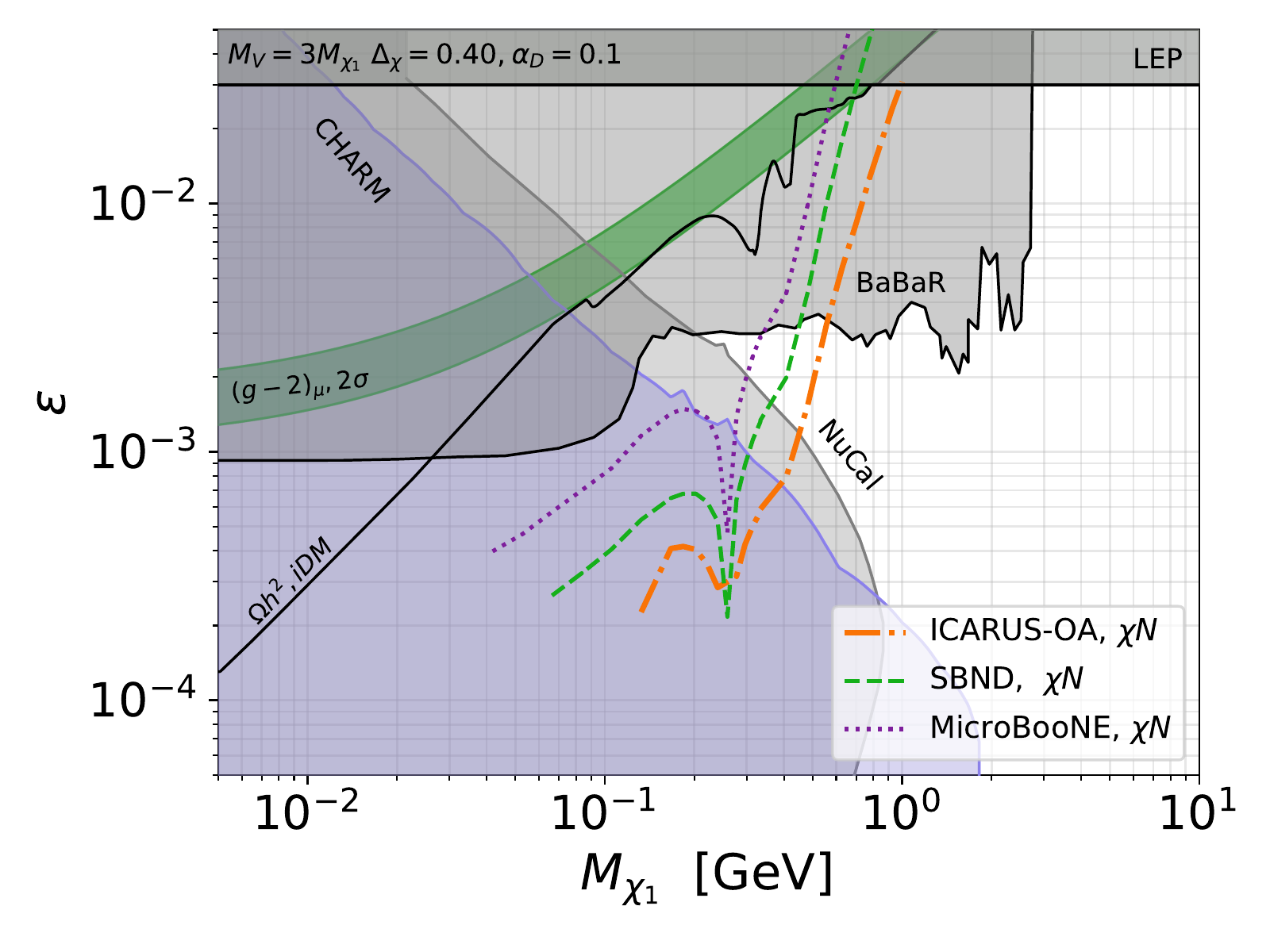}
	\includegraphics[width=0.99\linewidth]{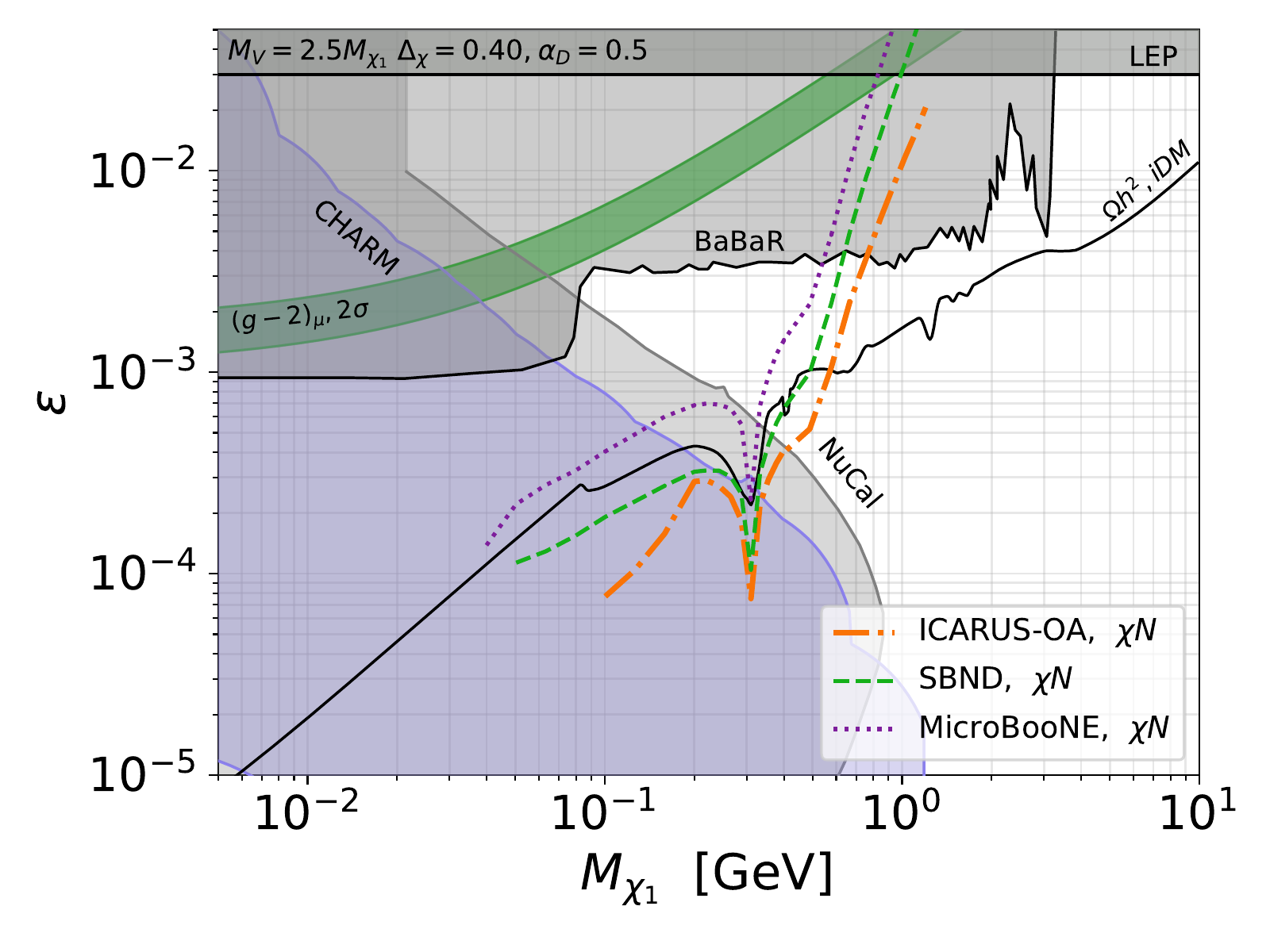}
	\hspace{0.cm}
\caption{Number of expected nucleon up-scattering events for the SBN experiments as function of the  mass $\mdm$ for $\dchi = 0.4$ with (top figure) $\alpha_D = 0.1, \Rx=3$ and (bottom figure)  $\alpha_D = 0.5, \Rx=2.5$. The dashed and dotted lines represent the kinetic mixing leading to $3$ events from $\chi N \to \chi N$ scattering, with $N$ a nucleon in the LAr target in the SBN experiments. We have estimated the CHARM  sensitivity  (grey purple region) at 95\% C.L. ($3$ events) results based on the procedure described in the main text. The upper grey region is excluded by DELPHI~\cite{Abdallah:2008aa,Fox:2011fx}, and BaBar~\cite{Lees:2017lec,Duerr:2019dmv}. The lower grey region is the NuCal exclusion from~\cite{Tsai:2019mtm}. The green region is the $2\sigma$ range for the $\gmtwomu$ anomaly~\cite{Abi:2021gix}.}
	\label{fig:vacuumexp}
\end{figure}

We illustrate the potential reach of the SBN experiments in the large mass splitting regime in Fig.~\ref{fig:vacuumexp}. Our projections assume the full anticipated datasets (see Table~\ref{tab:exp}): $6.6 \cdot  10^{20}$ PoT from the BNB for  MicroBooNE and SBND and $7.7 \cdot 10^{21}$ PoT from NuMI beam for ICARUS-OA.
For each experiment we display lines corresponding to $3$  $\chi_1$-nucleon up-scattering events,
assuming that the excited state $\chi_2$ subsequently decays in the detector, leading to a prompt or displaced $e^+e^-$ pair. 
In absence of dedicated background estimates for this particular signal channel, the results in Fig.~\ref{fig:vacuumexp} should be considered as qualitative possible targets for the SBN experiments in this range of the parameter space.
It is worth emphasising that for the relevant coupling ranges $\varepsilon \sim 10^{-3}$, the typical decay length times boost factor of $\chi_2$ is sub-metric, implying that most of the up-scattering events will be followed by a displaced decay in the detector, offering a clean way of further reducing the background. 
In view of this, we restrict our projections Fig.~\ref{fig:vacuumexp} to parameter regions corresponding to relatively short decay lengths, $c \tau_{\chi_2}$ smaller than around $10$ m.
We observe that all three SBN experiments can probe the thermal relic density target, which is indicated as a solid black line. 
In particular, for $\Rx = 2.5$, we  find that a significant part of the unconstrained parameter space predicting the observed relic abundance can be probed by the SBN experiments, corresponding to DM masses in the several hundred MeV range. 
Note that in the largest mass range, the state $\chi_2$ can also decay to final states containing a pair of muons.  
While beyond our current scope, the di-muon final state represents another interesting experimental signature that could be leveraged by the SBN experiments. 

Along with our SBN projections, Fig.~\ref{fig:vacuumexp} displays the existing constraints from DELPHI~\cite{Abdallah:2008aa,Fox:2011fx}, CHARM~\cite{Santoni:1986yw,Bergsma:1983rt,Bergsma:1985qz,Tsai:2019mtm}, NuCal~\cite{Barabash:1992uq,Blumlein:1990ay,Tsai:2019mtm}, and BaBar~\cite{Lees:2017lec,Mohlabeng:2019vrz,Duerr:2019dmv}, as discussed in Section~\ref{sec:otherlimits}. We have additionally derived a re-interpretation of the CHARM result including in particular the parton-level production not considered in~\cite{Bergsma:1983rt,Tsai:2019mtm}, significantly extending the reach to the larger mass range. Along with these existing constraints, there are several ongoing/proposed experiments which can probe the iDM model. We will defer our discussion of these experiments to the end of this section.

\begin{figure}[t]
	\includegraphics[width=0.99\linewidth]{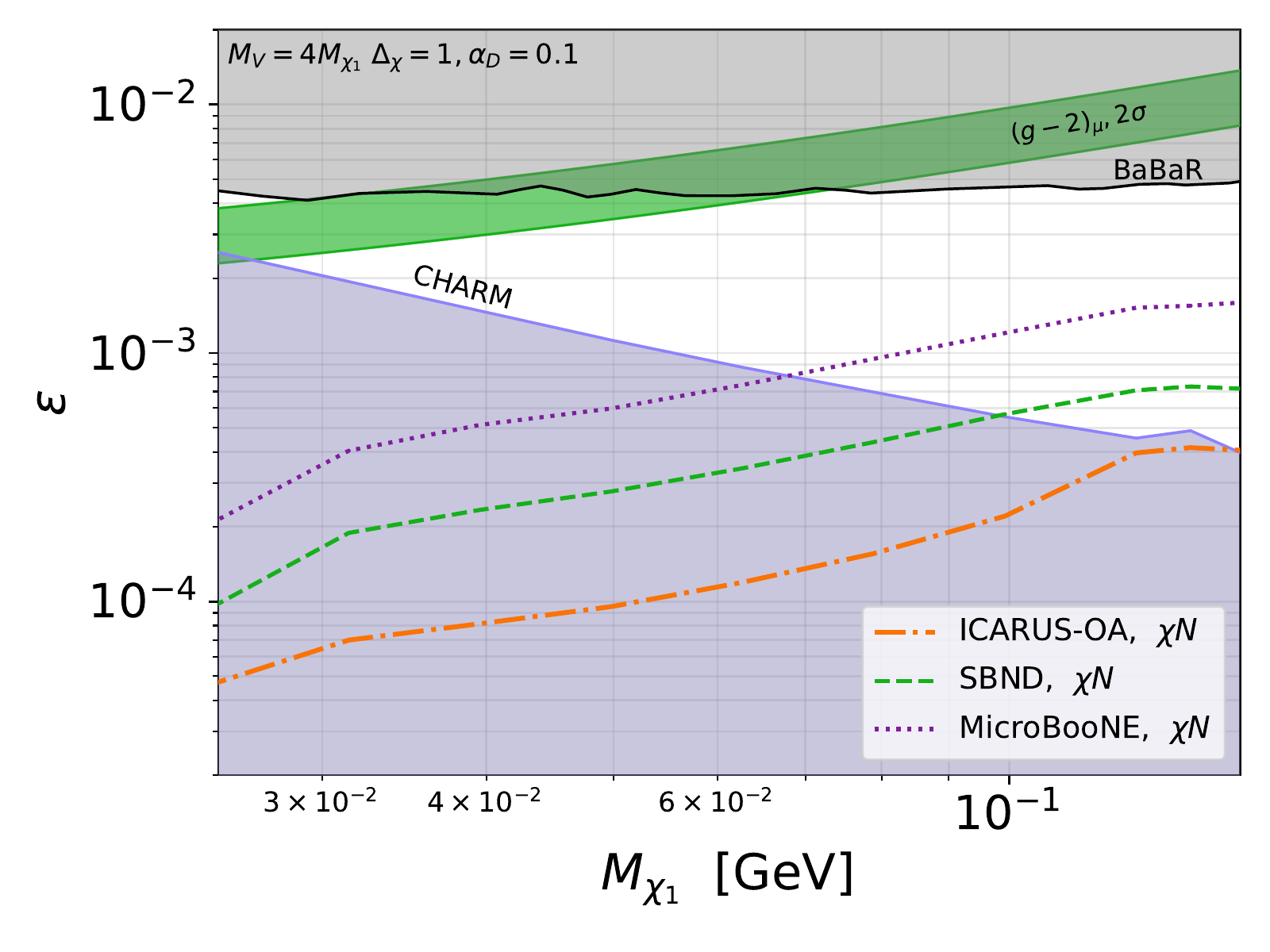}
	\hspace{0.cm}
\caption{Number of expected up-scattering events for the SBN experiments as function of the  mass $\mdm$ for $\dchi = 1.0$, $\alpha_D = 0.1$ and $\Rx = 4$. The dashed and dotted lines represent the kinetic mixing leading to $3$ events from $\chi N \to \chi N$ scattering, with $N$ a nucleon in the LAr target in the SBN experiments. We have estimated the CHARM  sensitivity  (grey purple region) at 95\% C.L. ($3$ events) based on the procedure described in the main text. The upper grey region is excluded by BaBar~\cite{Lees:2017lec,Duerr:2020muu}. The green region is the $2\sigma$ range for the $\gmtwomu$ anomaly~\cite{Abi:2021gix}.}
	\label{fig:LargeSplit4}
\end{figure}

Finally, we note that the large mass splitting region is particularly interesting since the short $\chi_2$ decay length may open a region of parameter space that is compatible with both an explanation of the $\gmtwomu$ anomaly~\cite{Abi:2021gix} and the relic density target, see e.g., Ref.~\cite{Mohlabeng:2019vrz}. While recasts of existing BaBar searches already probe the $\gmtwomu$ interpretations shown in Fig.~\ref{fig:vacuumexp}~\cite{Duerr:2019dmv,Kang:2021oes}, we find the SBN experiments can provide an independent dedicated tests of these important parameter regions. 
Furthermore, for very large splittings $\dchi \sim 1$, it was recently pointed out in~\cite{Duerr:2020muu} that a fraction of the $(g-2)_\mu$-favoured parameter space remains unconstrained by BaBar mono-photon searches. We illustrate in Fig.~\ref{fig:LargeSplit4} the potential sensitivity of SBN experiments by showing the 3-event lines for DM-nucleon up-scattering followed by a $\chi_2$ decay, with the parameter choice $\Rx=4, \alpha_D =0.1$ and $\dchi=1$. In the $(g-2)_\mu$-favoured range of kinetic mixing, we expect around $10^4-10^5$ such events depending on the experiment, significantly above any potential backgrounds\footnote{The complete model from~\cite{Duerr:2020muu} additionally includes a dark Higgs boson to provide an additional annihilation mechanism (see also the earlier work~\cite{Darme:2018jmx}). The presence of the latter does not significantly modify the signature described in the main text as long as its mass is larger than $\dchi \mdm $.}

\subsection{Small splitting region}

\begin{figure}[h!]
	\includegraphics[width=0.99\linewidth]{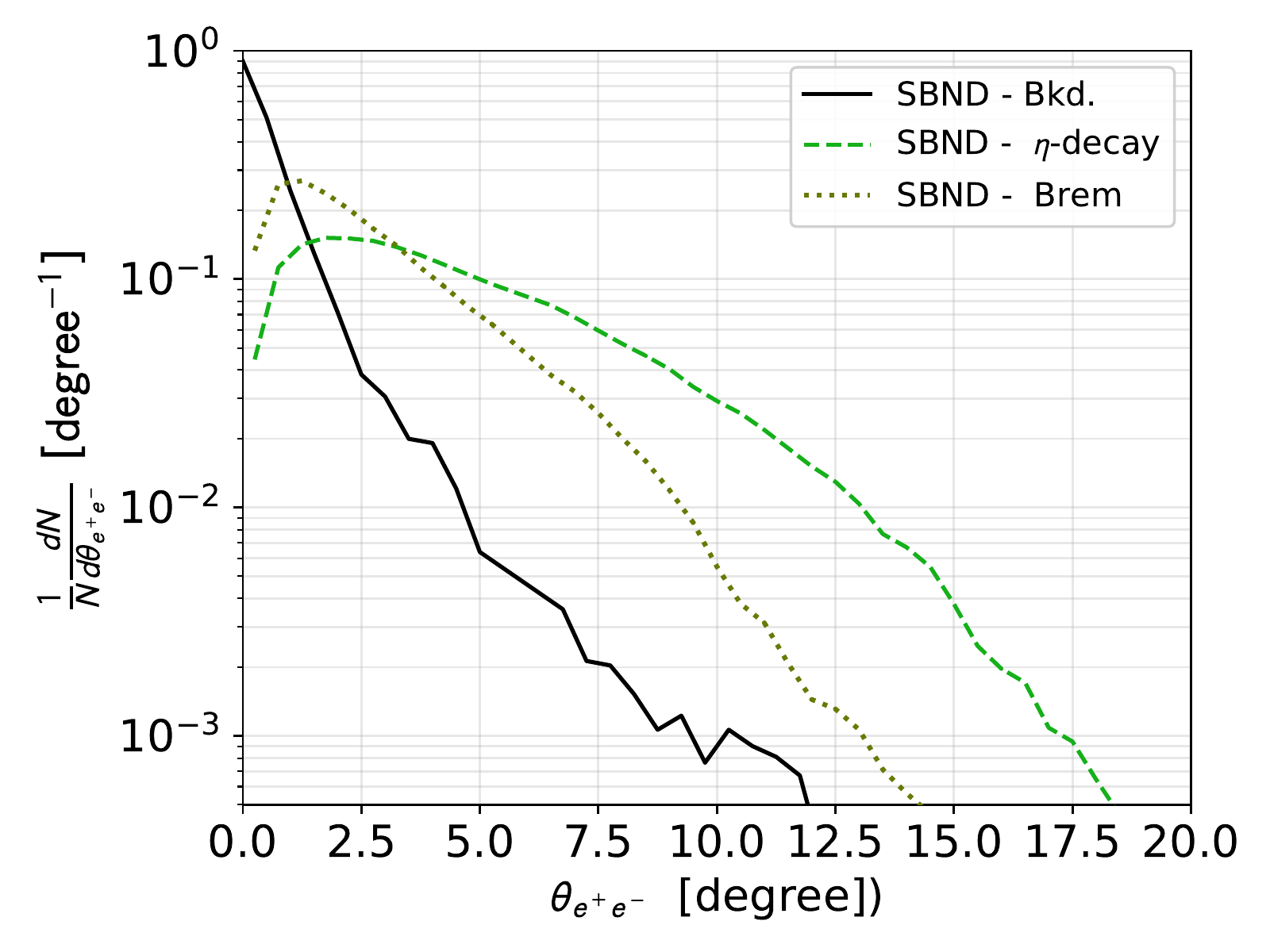}
	\hspace{0.cm}
\caption{ Normalised $e^+/e^-$ opening angle distribution of the signal and the neutrino-induced background for SBND for $\dchi = 0.1$ and $\mdm = 150$ MeV. The thick dashed line represents the background distribution. The solid thin line indicates the signal component originating from $\eta$ meson decay, while the dotted line represents the signal component from bremsstrahlung dark photon production.  }
	\label{fig:SigDistribution}
\end{figure}

We now turn to the second DM-relevant regime in which the mass splitting between the two states $\chi_2$ and $\chi_1$ is small, leading to long-lived or even quasi-stable $\chi_2$. We will first discuss the prospects for observing the semi-visible decays of beam-produced $\chi_2$ particles in the SBN detectors. As was noted in previous sections, the potential signals associated with such decays will have to compete with a mild neutrino-induced background.
In order to obtain accurate sensitivity projections, we include the background estimates described in the previous section, corresponding to around $20$k single photon background events for both the SBND and ICARUS-OA full datasets. As discussed in Sec.~\ref{sec:bkd} this background can be strongly reduced by selecting events with a large opening angle. Indeed, we illustrate in Fig.~\ref{fig:SigDistribution} the expected opening angle distribution of the final $e^+/e^-$ pair for both background and signal events for a DM mass of $\mdm = 150$ MeV at SBND. Signal events generically feature much larger opening angles than the dominant neutrino-induced background. We further show in Fig.~\ref{fig:SigDistribution} that the signal production mechanism, either meson decay or proton bremsstrahlung, plays an important role in fixing the shape of the signal distribution. The narrower opening angles predicted by proton  bremsstrahlung production reflect the fact that the dark photons originating from this process are typically much more energetic than their counterparts produced through meson decays.

\begin{figure}[h!]
	\includegraphics[width=0.99\linewidth]{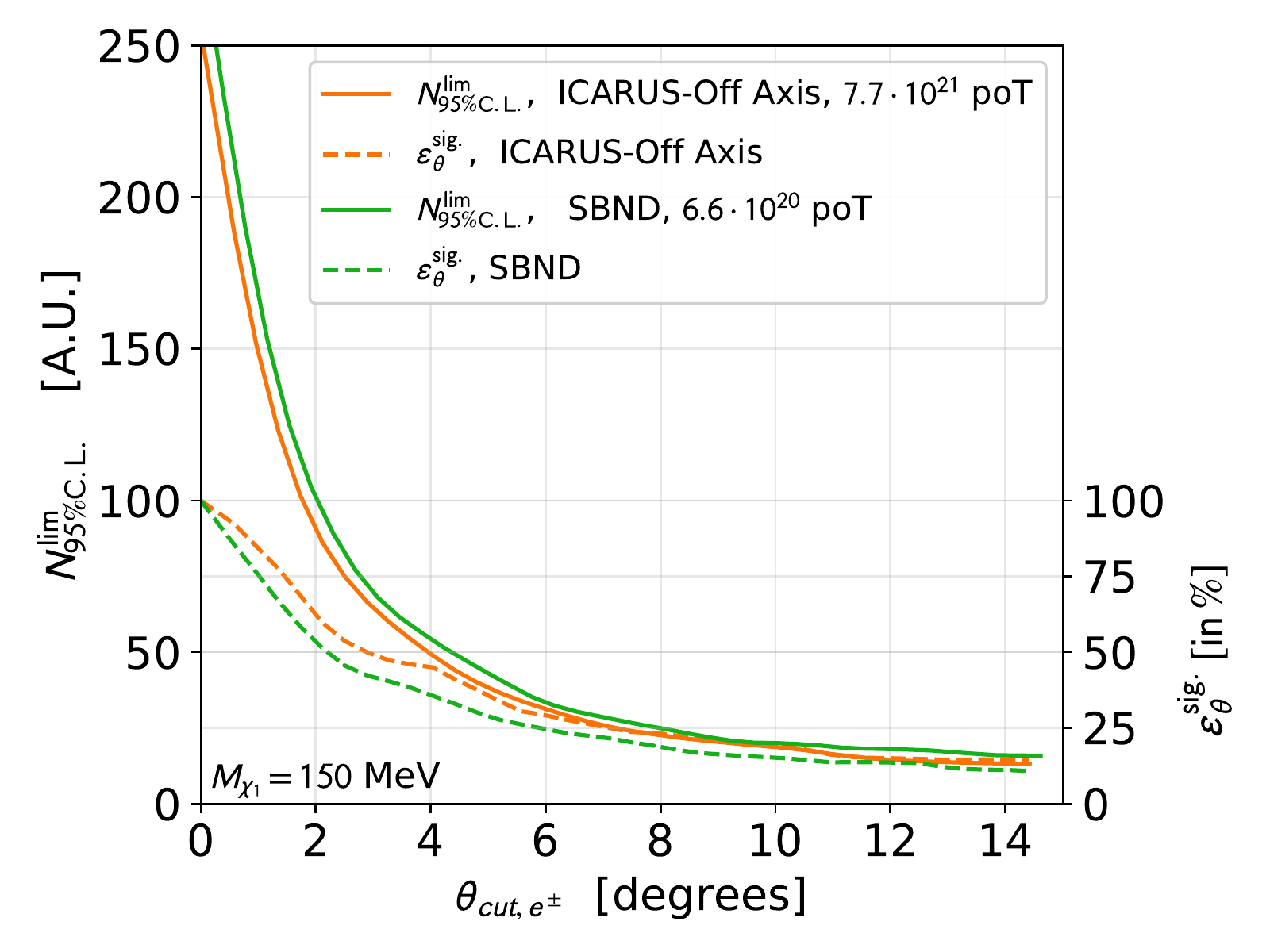}
	\hspace{0.cm}
\caption{Signal efficiencies of the opening angle cut 
($\epsilon^{\rm sig}_\theta$) for both ICARUS-OA 
(orange dashed line) and SBND (green dashed line), and the expected limit on the number of signal events at 95\% C.L. to overcome the neutrino-induced background in solid lines (orange for ICARUS-OA and green for SBND) for $\dchi = 0.1$ and $\mdm = 150$ MeV. }
	\label{fig:effangle}
\end{figure}

We estimate the usefulness of an opening angle selection cut by plotting in Fig.~\ref{fig:effangle} the efficiency of the cut for our signal sample, $\epsilon^{\rm sig}_\theta$, along with the expected 95\% C.L. limit on the number of signal events, $N^{\rm lim}_{ 95 \% \rm C.L.}$,  as a function of the opening angle cut, $\theta_{\rm cut}$.
It is clear that a $5^\circ $ cut can significantly reduce the background (as seen from the sharp decrease of the required $N^{\rm lim}_{ 95 \% \rm C.L.}$) at a moderate cost on the signal efficiency. Larger cuts do not lead to further improvement, but do not significantly reduce the signal to background ratio either. This indicates that the proposed search strategy may be viable even for experimental setups with only mild angular sensitivity. The main caveat is that the experimentally accessible mass range will be decreased for larger angular cut, which follows from Eq.~\eqref{eq:MassThreshold}. We will thus present our sensitivity estimates for long-lived $\chi_2$ semi-visible decays using a $\theta_{\rm cut} = 5^\circ $ cut. 

\begin{figure}[t!]
	\includegraphics[width=0.99\linewidth]{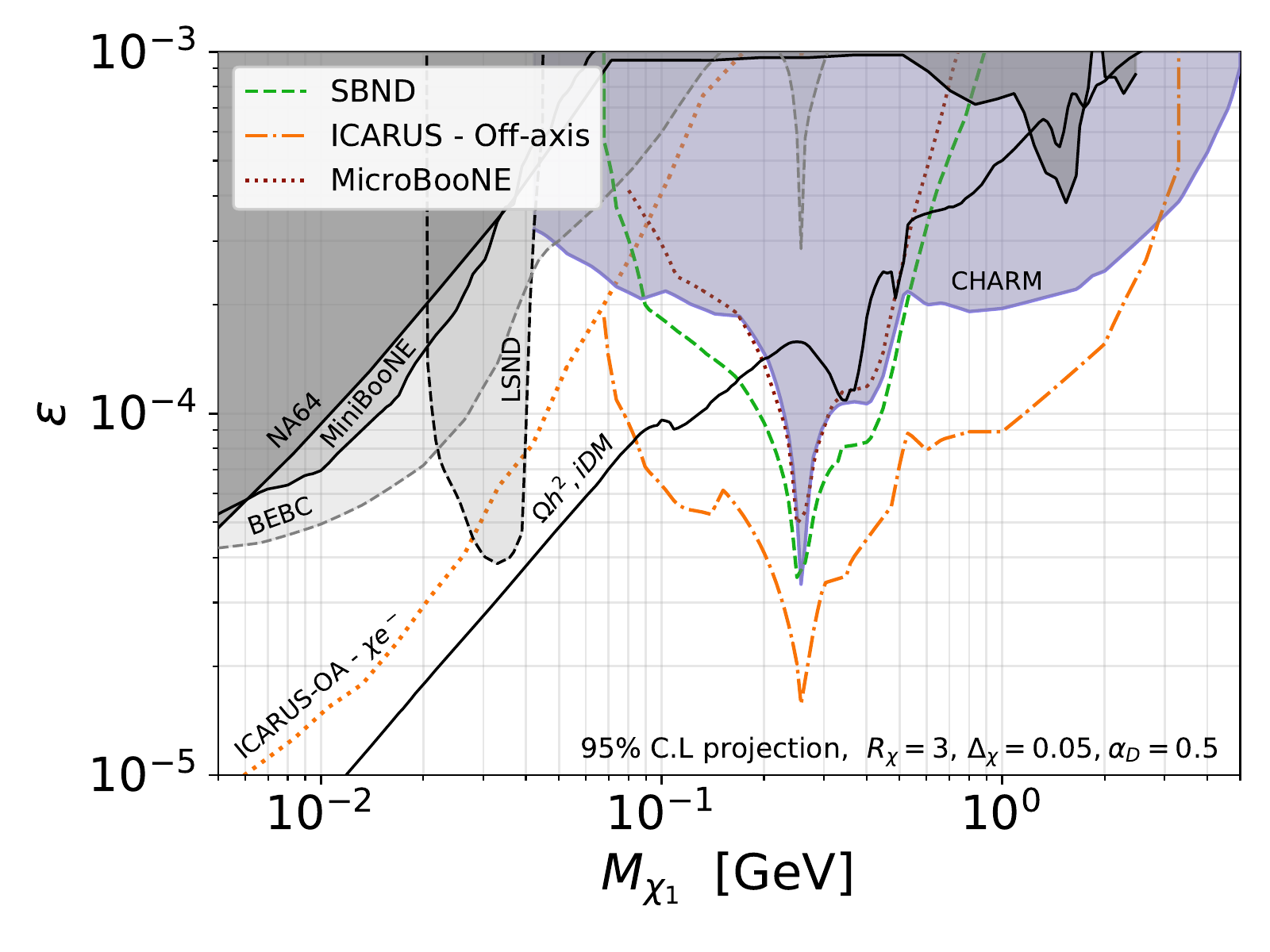}
	\includegraphics[width=0.99\linewidth]{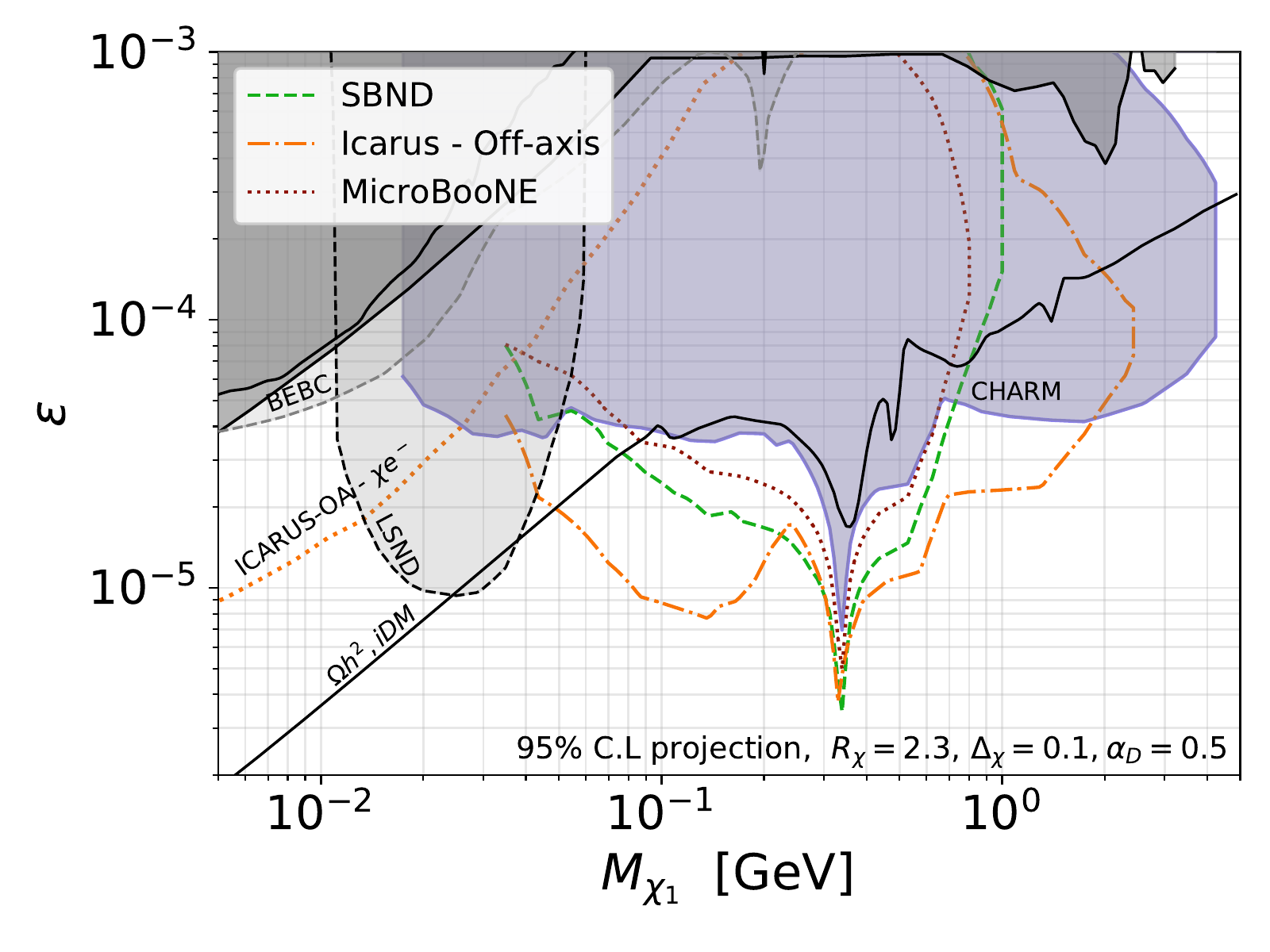}
	\caption{95\% C.L. limits for the SBN experiments for (top) $R_\chi = 3$, $\Delta_\chi=0.05$  (bottom)$R_\chi = 2.3$, $\Delta_\chi=0.1$. The grey region are excluded from NA64~\cite{NA64:2019imj} and BaBar~\cite{Lees:2017lec,Mohlabeng:2019vrz}. The LSND-excluded region at $\Delta_\chi=0.05$ is extracted from~\cite{Izaguirre:2017bqb} and recasted to $R_\chi = 2.3$, and directly extracted from~\cite{Tsai:2019mtm} for $\Delta_\chi=0.05$. The BEBC limit is extracted from~\cite{Grassler:1986vr,Buonocore:2019esg} and we included the MiniBooNE exclusion~\cite{Aguilar-Arevalo:2018wea}. The grey purple region is the CHARM bound the procedure from~\cite{Tsai:2019mtm}. The orange dot-dashed line is the reach of ICARUS in off-axis configuration, assuming $7.7 \! \cdot \! 10^{21}$ PoT from NuMI. The green dashed and purple dotted line are respectively the reach for SBND and MicroBooNE, assuming $6.6 \! \cdot \! 10^{20}$ PoT. Solid black line is the relic density target.}
	\label{fig:SBN0.05}
\end{figure}

Our results for long-lived $\chi_2$ decay signatures are presented in Fig.~\ref{fig:SBN0.05}, which shows the projected reaches for MicroBooNE, SBND, and ICARUS-OA based on their full expected datasets (see Table~\ref{tab:exp}). 

As discussed in the previous paragraph, we have imposed a cut on the $e^+/e^-$ opening angle, which helps to mitigate the neutrino-induced backgrounds. 
We observe that the SBN program can probe the relic density target (solid black line in Fig.~\ref{fig:SBN0.05}) for DM masses in the $50$-$200$ MeV range, where existing experimental constraints are not competitive. More broadly, the long-lived $\chi_2$ decay signatures can test a significant range of currently unexplored parameter space for DM masses between $50$ MeV and a few GeV. 

\begin{figure}[h!]
	\includegraphics[width=0.99\linewidth]{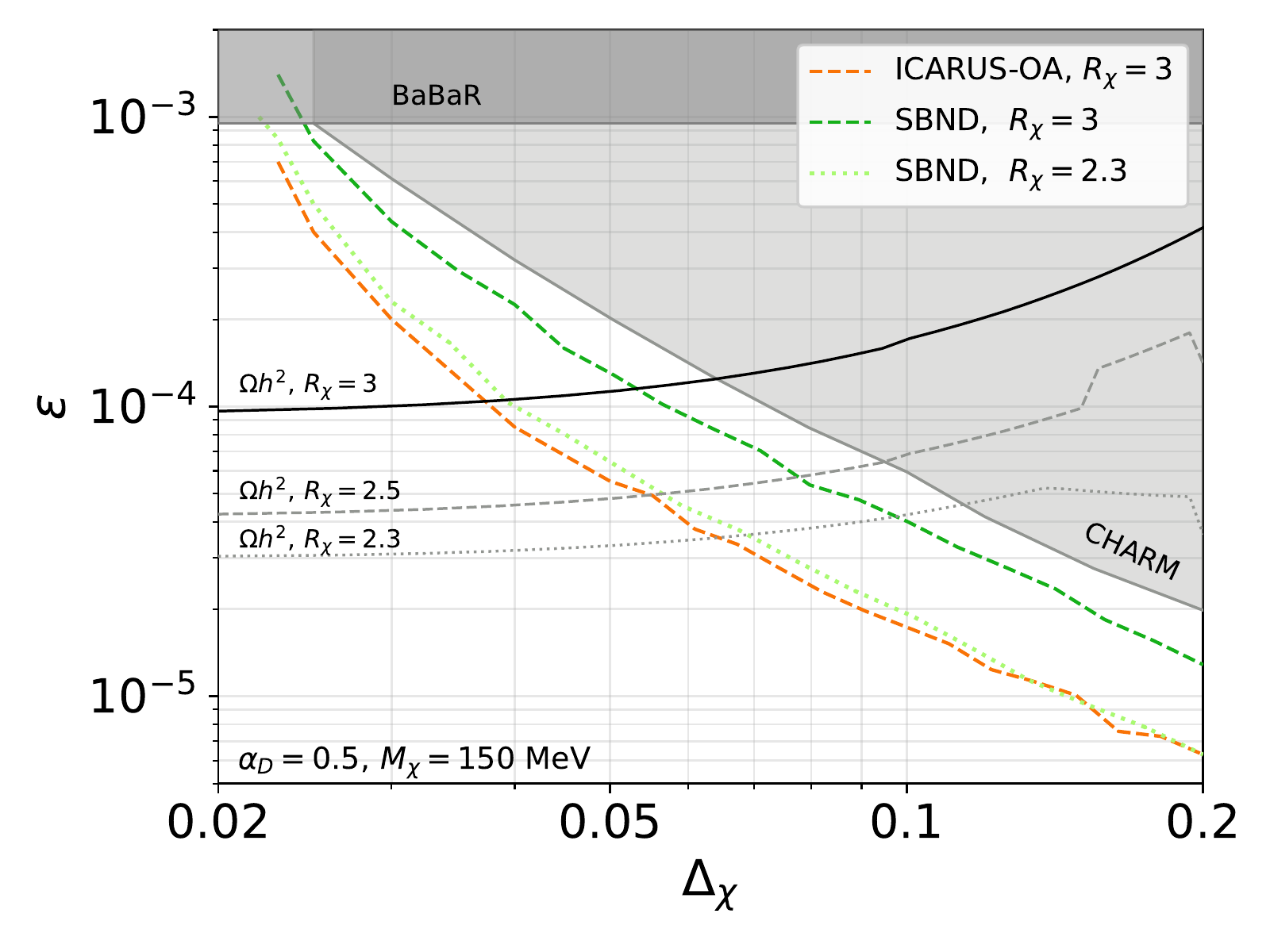}
	\caption{95\% C.L. limits for ICARUS-OA and SBND as function of the splitting between both  states $\chi_2$ and $\chi_1$. We took $\alpha_D = 0.5, \mdm = 150$ MeV. We show the target relic density lines for various choice of $R_\chi = \mdm/M_V$, note the limits are taken at $R_\chi = 3$ and will be slightly modified for smaller ratios.
}
	\label{fig:ICARUSdchi}
\end{figure}

As stressed above, although the angular cut on the electron/positron pair does not significantly modify the signal to background ratio for $\theta_{\rm cut} \gtrsim 5^\circ$, it directly impacts the lower mass threshold, following Eq.~\eqref{eq:MassThreshold}. For instance, increasing $\theta_{\rm cut} $ to $10^\circ$ would move the lowest accessible mass from $\sim 55$ MeV to $\sim 100 $ MeV in Fig.~\ref{fig:SBN0.05}. The mass threshold effect also reduces the reach of the SBN experiments as  $\dchi$ is lowered. We show this dependence in Fig.~\ref{fig:ICARUSdchi}, which displays the projected 95\% C.L. limits at SBN as a function of the mass splitting 
for a fixed DM mass $\mdm=150$ MeV. 
Although rapidly falling for smaller splitting, the SBN program will be able to probe a significant portion of the cosmologically motivated parameter space relevant for iDM.
 Since the experimental sensitivity for lower  masses is strongly limited by the kinematic threshold discussed in Sec.~\ref{sec:kinthre}, a ``low angular cut $\theta_{\rm cut}$ , large background'' approach could be an interesting additional search strategy. In particular, this could potentially provide independent coverage of the parameter space probed by the re-interpretation of LSND results, as discussed earlier in Sec.~\ref{sec:otherlimits}. Note that we quote the limits derived in~\cite{Tsai:2019mtm} for $\dchi = 0.05$ and from~\cite{Izaguirre:2017bqb} in $\dchi = 0.1$.

\begin{figure}[h!]
	\includegraphics[width=0.99\linewidth]{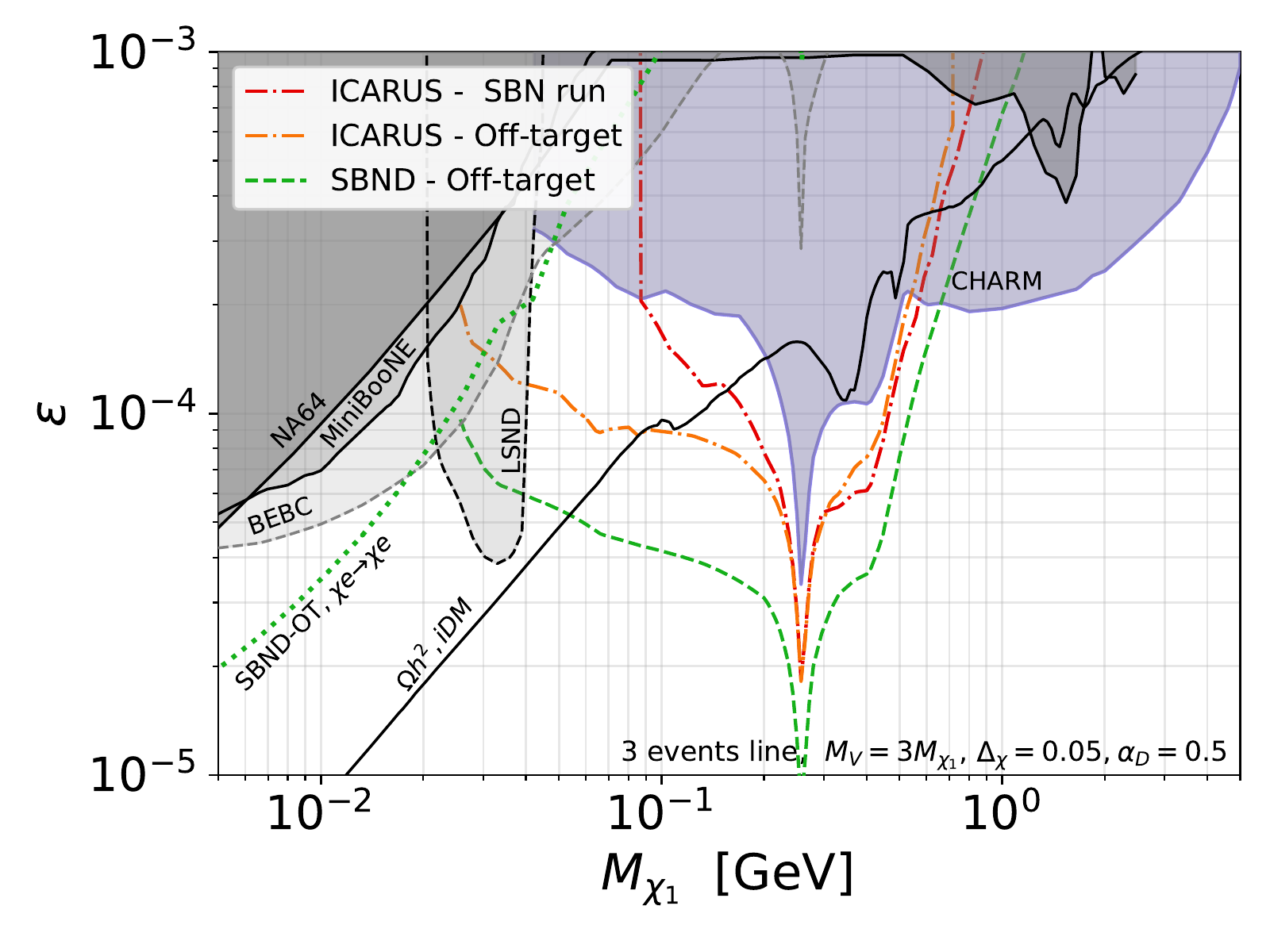}
	\caption{  3 events line for SBND (green dashed line) and ICARUS (orange dotted line) in an hypothetical ``off-target'' configuration, assuming $2.2 \! \cdot \! 10^{20}$ PoT. The red dashed line is the 3 events line for ICARUS during the standard SBN run (see Fig.~\ref{fig:SBN0.05}). The dotted green line represent the 3 $\chi e \to \chi e$ events line for SBND in off-target configuration. The grey region are excluded from NA64~\cite{NA64:2019imj} and BaBar~\cite{Lees:2017lec,Mohlabeng:2019vrz}. The LSND-excluded region is extracted from~\cite{Izaguirre:2017bqb}, the BEBC one from~\cite{Grassler:1986vr,Buonocore:2019esg} and we included the MiniBooNE limit~\cite{Aguilar-Arevalo:2018wea}. The grey purple region is the CHARM exclusion following the procedure from~\cite{Tsai:2019mtm}. Solid black line is the relic density target for $M_V = 3 M_{\chi_1}$, $\Delta_\chi=0.05$.}
	\label{fig:SBN0.05exotic}
\end{figure}

For small mass splittings and low DM masses in the few - tens of MeV range, the excited state $\chi_2$ will be quasi-stable, such that the decay signature is no longer effective. One can instead leverage $\chi_{1,2}$-electron scattering signatures (both up- and down-scattering) to provide coverage of this lower DM mass region.
Since the heavy state $\chi_2$ is very long-lived, the signature  closely mimics those of both the standard light DM elastic scattering and the familiar neutrino-electron scattering~\cite{Izaguirre:2017bqb}.
Following our discussion in Sec.~\ref{sec:bkd}, we require the scattered electron to be very forward with respect to the incoming $\chi_{1,2}$ direction, Eq.~(\ref{eq:CutsScatter}), which is expected to substantially suppress the dominant neutrino-electron scattering background for NuMI produced neutrinos at the off-axis ICARUS detector. We therefore present a $3$-events line for $\chi_{1,2}$-electron scattering in Fig.~\ref{fig:SBN0.05} for the ICARUS-OA setup, including the angular cuts on scattering signal given in Eq.~\eqref{eq:CutsScatter}.

Altogether, we find that the variety of distinctive signatures present in the iDM scenario -- long-lived $\chi_2$ decay, $\chi_1$-nucleon up-scattering followed by fast $\chi_2$ decay, and `standard' $\chi_
{1,2}$ up- and down- scattering off electrons -- probe complementary regions of parameter space and can significantly extend the reach beyond current experimental constraints.
We finally complement our study of the SBN program reach in standard neutrino (or anti-neutrino) run mode by showing in Fig.~\ref{fig:SBN0.05exotic} our projections for a potential off-target run configuration for both SBND and ICARUS, in which the proton beam is directed past the main beryllium target and onto the iron dump approximately 50 meters downstream.
We display 3-events lines for both long-lived $\chi_2$ decays and $\chi_{1,2}$-electron up- and down-scattering signatures. 
This off-target configuration was employed in a year-long run by the MiniBooNE collaboration in its dedicated light  search~\cite{Aguilar-Arevalo:2017mqx,Aguilar-Arevalo:2018wea}. Although we have only considered a $2.2 \! \cdot \! 10^{20}$ PoT dataset (analogous to the MiniBooNE off target run), the potential gain for the SBND collaboration is impressive. Indeed, such a setup could help dramatically in reducing the beam-related neutrino background rates due to the suppression of decay-in-flight neutrinos. Furthermore, the close proximity of the SBND experiment to the iron dump allows for a substantial increase in the angular acceptance of dark sector states compared to the standard Booster run mode.  

Besides the SBN program, there are number of existing and proposed experiments on the horizon that can probe the iDM model. In the near future, Belle-II~\cite{Kou:2018nap} will significantly improve on the BaBar limits, particularly with dedicated searches for displaced decays~\cite{Duerr:2019dmv,Duerr:2020muu,Kang:2021oes,Acevedo:2021wiq}. In the lower mass regime, NA64 will benefit from a substantial increase in statistics ($5\cdot 10^{12}$ electrons-on-target or more) in the coming years~\cite{Gninenko:2020hbd}, while next generation missing-energy/momentum experiments, such as LDMX~\cite{Akesson:2018vlm} or positron-based experiments~\cite{Battaglieri:2021rwp} have the potential of extending NA64 limits by more than an order of magnitude. 
Along with the SBN program, other proton beam fixed target experiments, such as DarkQuest/LongQuest~\cite{Berlin:2018pwi,Tsai:2019mtm}, JSNS$^2$\cite{Jordan:2018gcd}, and SHiP ~\cite{Anelli:2015pba} will have substantial sensitivity to iDM candidates in the sub-GeV range.
Finally, a rich experimental program is under development for long-lived particle detectors around the LHC interaction points~\cite{Chou:2016lxi,Feng:2017uoz, Gligorov:2017nwh,Curtin:2018mvb,Gligorov:2018vkc,Abreu:2019yak,Bauer:2019vqk,Ahdida:2020evc,Foroughi-Abari:2020qar,Batell:2021blf,Feng:2020fpf,Beacham:2019nyx}.
In tandem with the main general purpose detectors, the LHC can provide powerful and complementary sensitivity to light iDM~\cite{Izaguirre:2015zva,Berlin:2018jbm,Jodlowski:2019ycu}.

\section{Conclusions and outlook}
\label{sec:conclusion}

Proton beam fixed-target experiments, including accelerator neutrino beam facilities, offer a powerful means to search for light dark sectors. In this work we have examined the prospects of the three Fermilab SBN LArTPC detectors, MicroBooNE, SBND, ICARUS, to test models of light inelastic dark matter coupled to the SM through a kinetically mixed dark photon. 
Substantial fluxes of iDM states can be produced with both the on-axis 8 GeV Booster proton beam as well as the higher energy 120 GeV NuMi proton beam. Production from NuMI protons is particularly important for the large ICARUS detector, which is situated six degrees off-axis from the NuMI beamline. On the other hand, SBND, which is the detector closest to the BNB target, provides the best sensitivity to production from the Booster. 
Considering several experimental beam-target-detector run scenarios, we find that new substantial regions of iDM parameter space, including some predicting the observed thermal DM relic abundance, can be probed in the near future by all three experiments. 

The simple iDM model studied here leads to a rich phenomenology at the SBN experiments, furnishing several distinctive signatures including long-lived semi-visible decays of the excited dark state, DM-nucleon up-scattering plus fast de-excitation of the excited state, and up-and down-scattering with electrons. These signatures arise in qualitatively different regions of parameter space and thus offer complementary probes of the iDM model. 

The iDM signatures must be disentangled from the neutrino interactions the experiments were designed to detect. For the $e^+ e^-$ pair signature of the long-lived decay of the excited state, we have simulated the  dominant beam-related neutrino backgrounds for each of the experimental scenarios. Our findings suggest that a simple cut on the electron-positron pair opening angle can significantly mitigate the main background for this signal, which arises from conversion of secondary photons produced in neutrino interactions. Searches for long-lived decays of the excited state can probe new regions of parameter space, particularly for small mass splittings and DM masses in the 100 MeV range. 

For small mass splittings and low DM masses of order 10 MeV, where the excited state is effectively stable, up- and down-scattering leading to a very forward single electron can provide a relatively clean signal of the iDM model. On the other hand, for large splittings and larger DM masses of order 1 GeV, DM-nucleon up-scattering followed by prompt/displaced decays of the excited state in the SBN detectors offers a potentially striking probe. Though a full study of the potential neutrino-induced backgrounds is needed to accurately assess the SBN sensitivities, the large predicted signal event rates make this channel quite promising and worthy of further study by the collaborations. 

Within the proposed on-target run of the two beams, we highlight that further development of experimental techniques, such as tagging of soft baryons and exploitation of detailed timing information could help to further reduce the backgrounds considered in this work. Such possibilities merit further study as the understanding of LArTPC detectors improves. 

Although we focused primarily on the near-term prospects by estimating the experimental reach during the planned 3-year SBN run, we additionally explored the potential of a supplementary one-year run in an ``off-target'' configuration, in which the BNB proton beam is steered past the target onto the iron dump. While all experiments would benefit from the anticipated reduction of beam-related neutrino-backgrounds, the SBND detector was found to profit most from this setup, with almost an order of magnitude improvements in its potential reach. In this light, it would appear especially warranted to explore the technical and logistical feasibility, as well as other potential physics motivations, of such an off-target run.

The full SBN program is coming online and will reach maturity on the several-year time scale.  Our work reinforces the general expectation that these experiments have an excellent near-term opportunity to search for light new physics beyond the Standard Model, including light inelastic dark matter.\vspace{0.5cm}

\subsection*{ Acknowledgments}
\noindent
B.B. is supported by the U.S. Department of Energy under grant No. DE–SC0007914. 
L.D. is supported by the INFN ``Iniziativa Specifica'' Theoretical 
Astroparticle Physics (TAsP-LNF). 

\bibliographystyle{utphys}
\bibliography{biblio.bib}

\end{document}